\documentclass[preprint,journal]{vgtc}            

\onlineid{1669}

\ieeedoi{10.1109/TVCG.2023.3326603}

\vgtccategory{Research}

\vgtcpapertype{algorithm/technique}

\title{Extract and Characterize Hairpin Vortices in Turbulent Flows}

\author{%
  \authororcid{Adeel Zafar}{0009-0006-9997-5001},
  \authororcid{Di Yang}{0000-0002-4702-6393},
  \authororcid{Guoning Chen}{0000-0003-0581-6415}
}

\authorfooter{
  \item
  	Adeel Zafar, Di Yang, and Guoning Chen are with the University of Houston.
  	E-mail: azafar3@uh.edu, diyang@uh.edu, gchen16@uh.edu
}

\abstract{%

Hairpin vortices are one of the most important vortical structures in turbulent flows. Extracting and characterizing hairpin vortices provides useful insight into many behaviors in turbulent flows. However, hairpin vortices have complex configurations and might be entangled with other vortices, making their extraction difficult. In this work, we introduce a framework to extract and separate hairpin vortices in 
\revise{shear driven} turbulent flows for their study. Our method first extracts general vortical regions with a region-growing strategy based on certain vortex criteria (e.g., $\lambda_2$) and then separates those vortices with the help of progressive extraction of ($\lambda_2$) iso-surfaces in a top-down fashion. This leads to a hierarchical tree representing the spatial proximity and merging relation of vortices. After separating individual vortices, their shape and orientation information is extracted. Candidate hairpin vortices are identified based on their shape and orientation information as well as their physical characteristics. An interactive visualization system is developed to aid the exploration, classification, and analysis of hairpin vortices based on their geometric and physical attributes. We also present additional use cases of the proposed system for the analysis and study of general vortices \revise{in other types of flows}.
}


\graphicspath{{figs/}{figures/}{pictures/}{images/}{./}} 

\usepackage{tabu}                      
\usepackage{booktabs}                  

\usepackage{mathptmx}                  
\usepackage{float}
\usepackage{multicol}
\usepackage{tablefootnote}
\usepackage{multirow}
\usepackage{graphicx}

\usepackage{color, colortbl}
\definecolor{myred}{rgb}{1,0.3,0.3}

\definecolor{mygreen}{rgb}{0,0.8,0.3}

\newcommand{\revise}[1]{{\textcolor{black}{#1}}}

\begin{document}

\setlength{\baselineskip}{0.995 \baselineskip}

\setlength{\abovedisplayskip}{0pt}
\setlength{\belowdisplayskip}{0pt}
\setlength{\abovedisplayshortskip}{3pt}
\setlength{\belowdisplayshortskip}{2pt}
\setlength{\belowcaptionskip}{-3pt}
\newlength{\subfigbottomskip}
\addtolength{\subfigbottomskip}{-25pt}
\newlength{\subfigtopskip}
\addtolength{\subfigtopskip}{-25pt}
\setlength{\textfloatsep}{9pt}
\setlength{\floatsep}{3pt}
\setlength{\intextsep}{2pt}


\firstsection{Introduction}

\maketitle

\label{sec:intro}

Turbulent flow arises in applications ranging from automobile and aircraft engineering, climate study, and oil and gas engineering, to disease diagnosis and drug development. Understanding turbulent flows is a critical task for the success of these applications. \emph{Coherent structures} play a significant role in the understanding of various turbulent flow mechanisms, such as energy transfer and dissipation \cite{nguyen2020Taylor,nguyen2021DMD}.
Among all coherent structures, vortices are of particular interest to the experts in various turbulent flow studies and applications. 
Vortices are flow phenomena in which the flow particles move around some common lines or curves. Vortices are one of the most important dynamics in flow that often relate to energy/material transport and mixing \cite{wu2007vorticity}.
There exist numerous techniques for the extraction and visualization of vortices in different flows \cite{gunther2018state}. 
\revise{Due to the inherent complexity of turbulent dynamics and the multi-scale nature of coherent structures in turbulence, it is challenging to directly apply these techniques~\cite{bremer2015identifying}}.



\begin{figure}[t]
 \centering 
    \includegraphics[width=1\columnwidth]{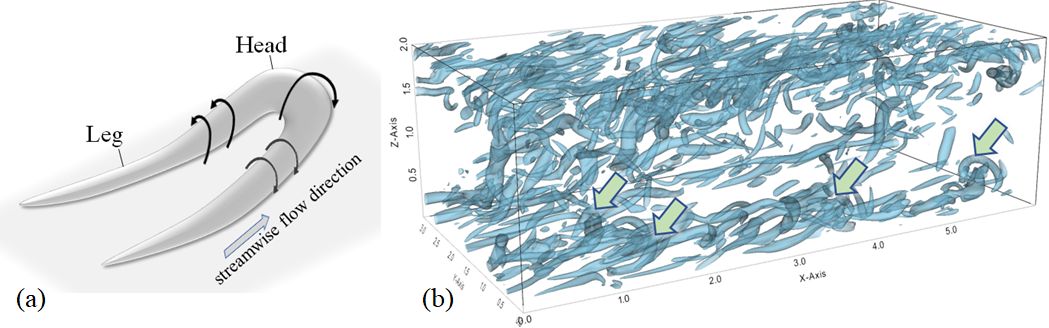}
    \vskip -1pt
    \subfigsCaption{(a) Illustration of a hairpin vortex, and (b) isosurfaces of $\lambda_2$ of a stress-driven turbulent Couette flow~\cite{li2019direct}. Arrows point to places where hairpin vortices may arise.}
 \label{fig:hairpinintro}
\end{figure}

Among different vortices arising in turbulent flows, hairpin vortices are of particular interest in the study of flow behaviors near the boundary layers. Hairpin vortices (\cref{fig:hairpinintro}) are formed due to turbulence in the vicinity of boundary layers in fluid flows~\cite{li2019direct}. 
Identifying and visualizing hairpin vortices provides crucial insight into the transition process of fluid flows from laminar to turbulent around the fluid boundary layers~\cite{adrian2007hairpin}. \revise{Hairpin vortices are also important in other phenomena such as material transport from the low speed fluid close to the boundary layer towards the high speed fluid away from the boundary layer (e.g., the lifting of  dust from the ground for the formation of haze).}
However, in practice, boundary layers in turbulent flows have complex configurations, resulting in hairpin vortices in irregular shapes and varying sizes (\cref{sec:hairpin}). They are also often tangled with other vortices, making their extraction difficult.  
Existing methods that rely on thresholding strategies applied to certain physical attributes usually lead to incomplete (or disconnected) vortices due to the sensitivity of the selection of a proper threshold value. There is currently a lack of a robust and automatic framework to extract hairpin vortices. 

To address these challenges, we develop a new and robust framework for hairpin vortex extraction and characterization in \revise{shear driven turbulent flows (e.g. channel, couette, wake and pipe flow)}. In particular, our framework makes the following contributions.

\begin{itemize}
  \setlength{\itemsep}{0pt}
  \setlength{\parskip}{0pt}
    \item We introduce a region-growing process, followed by a region-splitting process for vortex detection and separation (\cref{sec:vortexextraction}) This aids the selection of a global threshold for vortex detection and yields a simpler tree than the conventional contour tree, representing the spatial hierarchy relationships among vortices.
    \item We build a profile for each vortex based on its physical and geometric properties (\cref{sec:vortexprofiling}), which can be used for the classification, statistical analysis, and interactive exploration of individual vortices. To extract the shape and orientation information of individual vortices, we adapt the recent skeletonization technique from the geometry processing community.
    \item We define the first automatic pipeline to separate candidate hairpin vortices from all vortices based on experts' knowledge of hairpin vortices (\cref{sec:hairpin}). These candidate hairpin vortices contain all important hairpin vortices. A subsequent clustering of them separates them from other non-hairpin vortices.
    \item We devise an interactive visualization system (\cref{sec:visualizationsystem}), allowing the users to explore and inspect the extracted vortices in both the physical space and the attribute spaces. The users can not only inspect the detailed characteristics of the individual vortices and their relations with nearby vortices but also select groups of vortices for study.
\end{itemize}

We have applied our method and the visualization system to a number of flows to evaluate their effectiveness. In particular, we apply our method to aid the extraction and study of hairpin vortices in the stress-driven turbulent Couette flow \cite{li2019direct} (\cref{sec:results}). Our results show that our system can robustly identify hairpin vortices with different shapes, corresponding to different stages of the hairpin vortices, that are hard to identify with the existing methods. \revise{Furthermore, our system can also be applied to the study of other types of vortices in flows with predominant streamwise direction. To demonstrate this, we present few use-cases for several small scale flows}.


\section{Related Work}
\label{sec:related_work}

\revise{Vortex extraction and analysis continue to be active areas of research in fluid dynamics and scientific visualization~\cite{post:state,TimeDepTopoSTAR11,heine2016survey,gunther2018state}. Vortex definitions~\cite{gunther2018state} commonly involve two key components: the vortex coreline, which represents the path around which fluid particles move~\cite{lugt1979dilemma}, and a reference frame that reveals circular or spiral patterns when streamlines are projected onto a plane perpendicular to the coreline~\cite{robinson1991coherent}. Different methods have been developed to identify vortex corelines and regions.
For coreline extraction, \emph{line-based methods} ~\cite{BAN94,sujudi1995identification,PEI99,weinkauf07c} such as the Parallel Vector operator~\cite{PEI99} and the reduced velocity criterion~\cite{weinkauf2010streak} are widely used. However, these methods often result in numerically unstable and fragmented corelines, posing challenges for accurate extraction. \emph{Region-based methods}, on the other hand, rely on scalar quantities like pressure~\cite{hunt1988eddies}, vorticity, and various criteria (such as Q~\cite{hunt1987vorticity}, $\Delta$~\cite{chong1990general} and $\lambda_2$~\cite{jeong1995identification} criterion) to characterize rotation behavior locally. These methods typically require threshold values, which can impact the size and extent of the extracted vortices. To address this issue, topological segmentation approaches~\cite{bremer2015identifying, schneider2008interactive} have been introduced to identify vortices in turbulent flows without the need for a global threshold.
In addition to local methods, global approaches have been developed to construct the skeleton of the vortex tube. \emph{Geometric-based methods}, including curvature center method~\cite{sadarjoen2000detection} and winding angle method~\cite{sadarjoen2000detection}, leverage streamline shapes to identify vortices. However, certain classes of vortices, such as those moving along non-linear paths, may not be extracted effectively using these approaches. \emph{Integration-based methods}, such as particle density estimation~\cite{wiebel2011topological} and Jacobian analysis~\cite{weinkauf2010streak}, offer alternative solutions by observing the attraction behavior of injected particles over time. Vortex behavior can also be studied through physical attribute characteristics~\cite{nguyen2021physics} and pairwise correlations~\cite{berenjkoub2018visual} along individual pathlines. \emph{Objective methods}~\cite{haller2005objective,haller2015lagrangian} have gained attention, aiming to establish a steady reference frame in which vortex behavior can be objectively identified~\cite{gunther2017generic,hadwiger2019time,BaezaRojo19}. One such objective characterization~\cite{haller2015lagrangian} is the largest nested elliptic LCS, which represents a material line preserving arc length in incompressible flow. Similarly, Salzbrunn et al.~\cite{salzbrunn2006streamline, salzbrunn2008pathline} introduced streamline predicates and pathline predicates, respectively, to extract vortices and flow structures.}

Machine learning (ML) techniques have also been applied to extract vortices. 
Deng et al.~\cite{deng2019cnn,wang2020rapid} applied supervised training for vortex extraction. They derived a vortex ground truth from the velocity field by applying a user-defined threshold to the instantaneous vorticity deviation (IVD)~\cite{haller2016IVD} in order to produce a binary mask, identifying vortices and non-vortices. 
Given the sea surface height, Lguensat et al.~\cite{lguensat2018eddynet} extracted ocean eddies.
Franz et al.~\cite{franz2018ocean} detected ocean eddies by training a neural network that receives a vortex measure, e.g., the Okubo-Weiss criterion as input~\cite{okubo1970horizontal,weiss1991dynamics}. To track the vortices over time, they applied a recurrent neural network (RNN) afterward.
Bai et al.~\cite{Bai19:OceanEddyCNN} sent images of streamlines into a CNN to detect ocean eddies.
Kim and G{\"u}nther~\cite{Kim19EuroVis} developed a CNN that extracts a reference frame in which an unsteady flow becomes steady, enabling vortex coreline extraction.
Berenjkoub et al. \cite{berenjkoub2020vortex} produced synthetic flows with 2D vortices of various shapes for the training of different neural networks to aid the extraction of vortex boundaries.

While the above methods may extract individual vortices to some extent, their application to the turbulent flow that contains vortices with multiple different scales remains a challenging task. To partially address that, Nguyen et al. \cite{nguyen2020Taylor} introduced a visualization framework to separate the large-scale Taylor vortices from the small-scale vortices for the study of the Taylor-Couette turbulent flow. Later, they proposed to use the dynamic mode decomposition (DMD) for the separation of large-scale coherent structures from the smaller ones in the turbulent shear flow \cite{nguyen2021DMD}. Nevertheless, their techniques focus mostly on the extraction of the largest structures in the flow. 
\revise{Moreover, there have been studies to detect other flow structures in near-wall turbulence. For example, Nsonga et al. in ~\cite{nsonga2019detection} introduced methods for the detection and visualization of splat and anti-splat events in turbulent flows, contributing to the understanding of these phenomena. Also Nsonga et al. in ~\cite{nsonga2019analysis} focused specifically on the analysis of near-wall flow in a turbine cascade using splat visualization techniques, offering valuable insights into the flow characteristics in this particular scenario.}




In summary, there does not exist a comprehensive framework specifically for the extraction of hairpin vortices in turbulent flows.

\begin{figure*}[!t]
 \centering 
 \includegraphics[width=0.9\linewidth, height=0.5\linewidth]{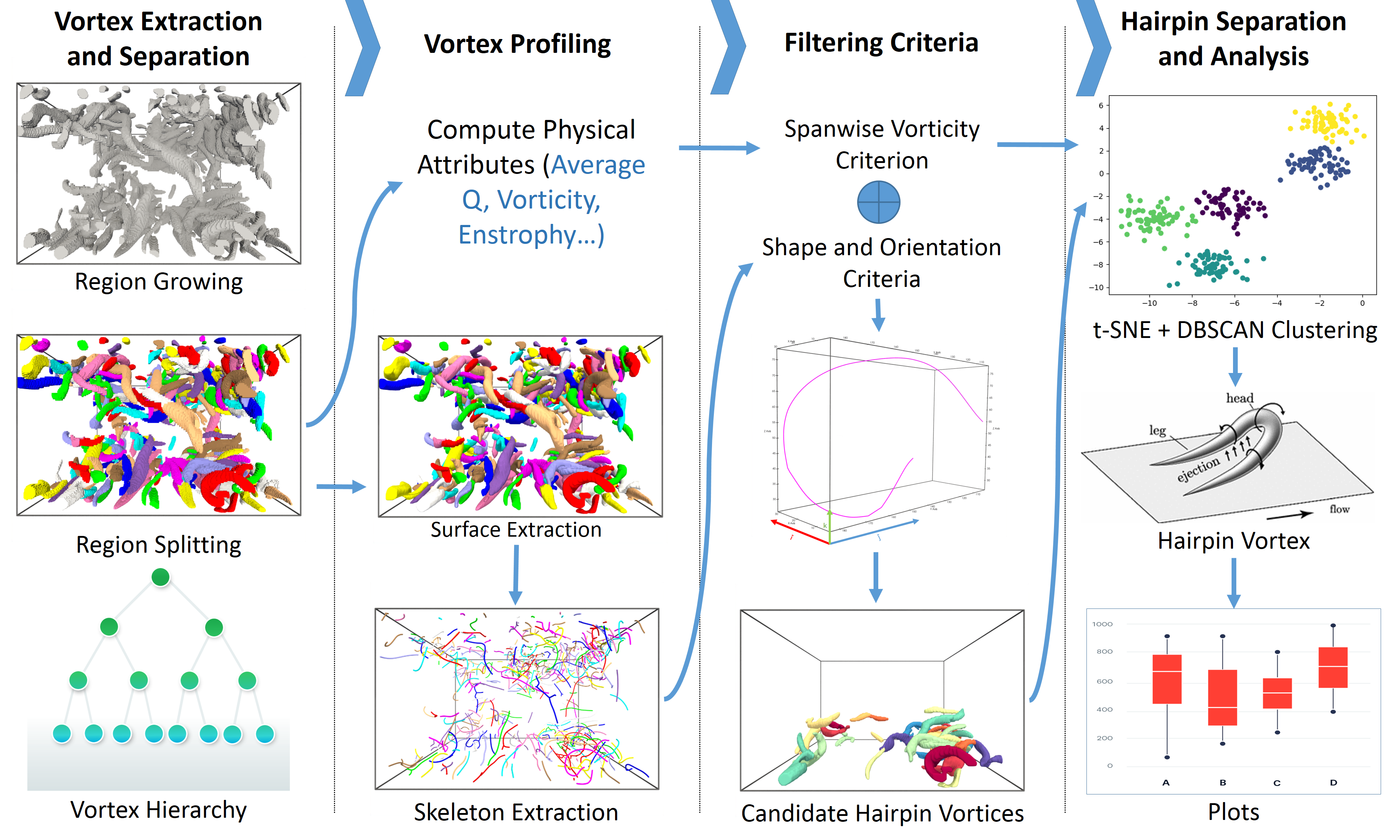}
 \subfigsCaption{The pipeline of our framework for the extraction and analysis of hairpin vortices. }
 \label{fig:fig2.1}
\end{figure*}

\section{Overview of Our Method}
We propose a new framework to address the challenges of extracting and separating vortices, especially hairpin vortices, of different scales for their analysis. Our framework consists of the following steps (\cref{fig:fig2.1}).

First, we identify the vortical regions using a region-growing process based on a selected vortical measure (e.g., $\lambda_2$). The goal of this region-growing is to identify regions that may contain vortices. Our region-growing strategy extends the seminal work by Banks and Singer~\cite{BAN94}.
The resulting regions are large and contain multi-vortices. Next, we perform a top-down region-splitting process to procedurally shrink the initial regions into disjoint sub-regions that contain single vortices. This results in a hierarchical spatial relation among vortices that can be represented as a tree. This hierarchical tree is similar to the well-known contour tree or split tree \cite{van1997contour,bremer2015identifying}. However, the nodes in our \revise{tree do not need to correspond to} the topological features (e.g., extrema and saddles) of the contour tree. We will explain why we choose our strategy over contour tree construction later.

Second, after obtaining the individual vortices, we construct their respective profiles. Each vortex profile consists of the physical attribute information (e.g., $Q$, acceleration, vorticity, etc.) and the geometric characteristics (e.g., shape, orientation, and size) of the corresponding vortex. The vortex profiles will be used for the subsequent classification of the types of vortices and their statistical study. We include the most well-known attributes and properties used for vortex identification and characterization in the literature to provide a more complete depiction of the vortical behaviors.

Third, to identify hairpin vortices, specifically, we set criteria based on discussion with the experts. The criteria allow us to select candidate vortices from the above initial set of vortices through a filtering process. The goal is to include all hairpin vortices. Other criteria can be defined to identify other types of vortices.

Finally, we identify the ``true'' hairpin vortices from the above candidates.
We use two strategies here. On the one hand, we use a simple clustering technique to separate the ``true'' hairpin vortices from the rest. On the other hand, we present an interactive visualization system to aid a manual selection of those hairpin vortices from the candidates. This system is also equipped with functionality to aid the statistical study of sets of hairpin vortices (and other general vortices).

In the following sections, we provide more details of the above steps.

\section{Vortex Extraction and Separation}
\label{sec:vortexextraction}

In this section, we will describe our region-growing strategy for the extraction of vortical regions followed by a top-down region-splitting process for the separation of vortices. 


\subsection{Vortical Region Extraction}
\label{section3.1}
According to the $\lambda_2$ criterion \cite{jeong1995identification}, the vortices are the regions where $\lambda_2 < 0$. However, $\lambda_2 < 0$ is less restricting and an appropriate threshold is needed to extract valid vortices\cite{chakraborty2005relationships}. As the range of $\lambda_2$ is dataset dependent, a common threshold is not feasible to extract vortices in different datasets. In order to get the dataset appropriate $\lambda_2$ value for vortex extraction (which we call $\lambda_{2i}$), we use a histogram refinement approach which consists of the following steps.
\begin{itemize}
  \setlength{\itemsep}{0pt}
  \setlength{\parskip}{0pt}
\item[i.] Compute a histogram of $n$ bins using the range [\textit{min}($\lambda_2$), \textit{max}($\lambda_2$)], where \textit{min}($\lambda_2$) and \textit{max}($\lambda_2$) are the minimum and maximum $\lambda_2$ values, respectively, \revise{considering only $\lambda_2 < 0$}. In practice, we set $n=100$ based on our experiments.
\item[ii.] Ignore all the bins that contain less than 0.1\% of the total voxels and compute a new histogram of $n$ bins with the new range [\textit{min}(\textit{hist}($\lambda_2$)), \textit{max}(\textit{hist}($\lambda_2$))] where \textit{min}(\textit{hist}($\lambda_2$)) and \textit{max}(\textit{hist}($\lambda_2$)) are minimum and maximum $\lambda_2$ values in the filtered histogram.
\item[iii.] Repeat steps i and ii until \textit{(a)} the last bin contains less than 30\% of voxels, \textit{(b)} the difference between the percentages of the last and second last bin is less than 20, \textit{(c)} if the last bin count in the new and the previous histogram is the same.
\end{itemize}

Step (ii) filters out less important values, and the stop conditions \textit{(a)} and \textit{(b)} make sure that the histogram is not refined too much. The thresholds are chosen after careful analysis of multiple datasets mentioned in \cref{tab:Table2}. $\lambda_{2i}$ is assigned the lower value of the 90th bin of the final histogram. \revise{The choice of parameters in the steps above affects the extent of the extracted vortices. Increasing the number of bins in step (i) and the thresholds for stop conditions \textit{(a)}, \textit{(b)} in step (iii) push the smaller $\lambda_2$ values towards the bins with the higher counts, then the $\lambda_{2i}$ value at the 90th bin will correspond to a smaller $\lambda_2$ value which will reduce the extent of the extracted vortices. Conversely, increasing the cut-off threshold of 0.1\% in step (ii) ignores more bins from the lower side of the histogram resulting in larger $\lambda_2$ values pushed towards the lower bins and consequently larger $\lambda_{2i}$.}

Given $\lambda_{2i}$, we find cells with local minimum $\lambda_2$ in the dataset. We convolve a window of size $3\times3\times3$ and find the cells with minimum $\lambda_2$ value locally. The local cell which doesn't have any point with $\lambda_2 < \lambda_{2i}$ is not considered for local minimum calculation. After this process, we end up with a list of local minimum cells. The local minimum cells are used as seeds for region growing.

In our region-growing strategy, we extract the regions using geometric connectivity and scalar criterion. For each seed cell, a neighboring cell is considered connected to the seed cell if one of its points fulfills the criterion that $\lambda_2 < \lambda_{2i}$. We then iteratively add new cells by checking the neighboring cells for the $\lambda_2 < \lambda_{2i}$ criterion. The region-growing stops when no further cells fulfill the criterion. 
The seed cells that overlap the already-grown regions are ignored. The regions which are smaller than a particular number of cells are considered noise and are ignored. In our implementation, we set this threshold as $0.01\%$ of the total number of voxels for each data set. This results in the extraction of vortical regions in the input flow. One such example of the extracted region is shown in \cref{fig:fig3.2.3a}. Note that, criteria other than $\lambda_2$ (e.g., $Q$-criterion, vorticity magnitude or pressure~\cite{BAN94, bremer2015identifying}) can be applied to guide the above region growing process. 

\subsection{Vortex Separation}
\label{section3.2}
We make use of the contour tree \cite{carr2003computing} to separate common regions of multiple vortices into their respective sub-regions. Given a point set $\{x \in \mathbf{R}^n\}$ and a scalar field $\{f(x) \in \mathbf{R}\}$, in the context of contour tree, a level set is the set of points $p$ where $f(p) = c$. When $n = 3$, the level set is an isosurface and $c$ is called an isovalue or contour value. As we change the contour value, the isosurface evolves (splits or merges). The contour tree represents the hierarchical relation between merging or splitting of the isosurfaces at different contour values. Traditional contour tree \cite{carr2003computing} uses topological critical points to change the level set. In 3D, it means a new isosurface is extracted using the scalar value at the critical point in the grid. At this stage, the isosurface may get split and the process continues iteratively on each isosurface component until no isosurface remains. This results in a dense graph/tree representing the evolution of isosurfaces in accordance with isovalues. We use $\lambda_2$ as the scalar field to extract the contour tree with some differences. The traditional contour tree is very dense and may suffer from degenerate splits because of the possible existence of a large number of critical points \cite{michalak2018realization}. In order to keep the tree as sparse as possible and to limit the degenerate splits, we use the histogram expansion strategy to pick $\lambda_2$ isovalues for the contour tree construction. The histogram expansion strategy works as follows.

\begin{itemize}
  \setlength{\itemsep}{0pt}
  \setlength{\parskip}{0pt}
\item[i] Set $n = 100$. Compute a histogram with $n$ bins in the range [\textit{min}($\lambda_2$), $\lambda_{2i}$], where \textit{min}($\lambda_2$) is the minimum $\lambda_2$ value in the dataset and $\lambda_{2i}$ is the initial $\lambda_{2i}$ value computed in \cref{section3.1}.
\item[ii] Stop, if the percentage of voxels in the final bin is less than 10\%, otherwise $n=2n$ and repeat step (i).
\item[iii] Sort the bins of the final histogram in descending order of $\lambda_2$.
\item[iv] Compute a Fibonacci series between $[0, n]$ and pick all the bins at the indices in the series from the sorted histogram. A list is computed consisting of the $\lambda_2$ values at lower values of the bins\footnote[1]{Here bin is a tuple representing a range of $\lambda_2$ values}. We call this list the $\lambda_2\_steps$.
\end{itemize}

\begin{figure}[t]
 \centering 
      \begin{subfigure}[b]{0.49\linewidth}
         \centering
         \includegraphics[width=0.99\linewidth]
         {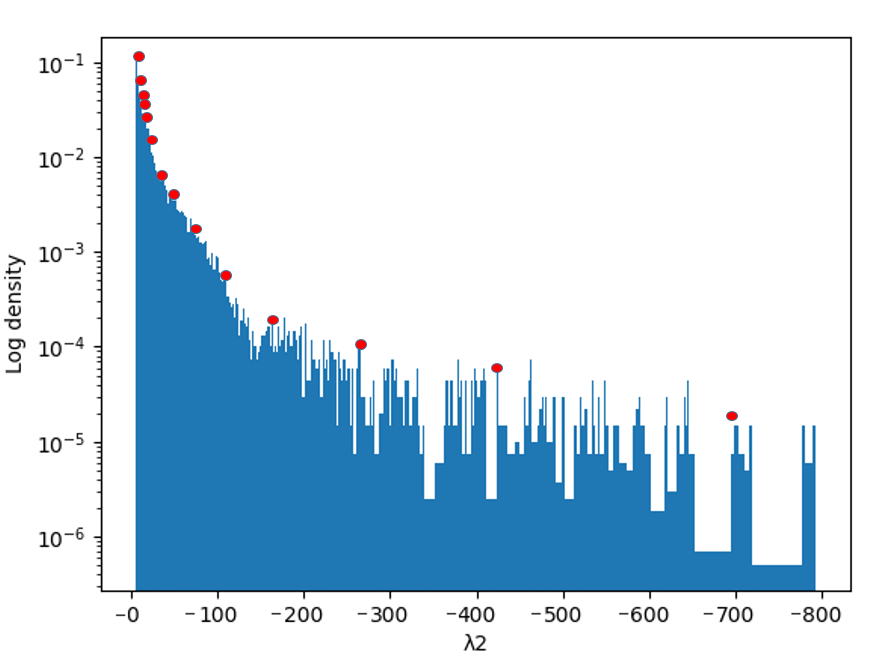}
         \vskip -4pt
         \caption{}
     \end{subfigure}
           \begin{subfigure}[b]{0.49\linewidth}
         \centering
         \includegraphics[width=0.99\linewidth]{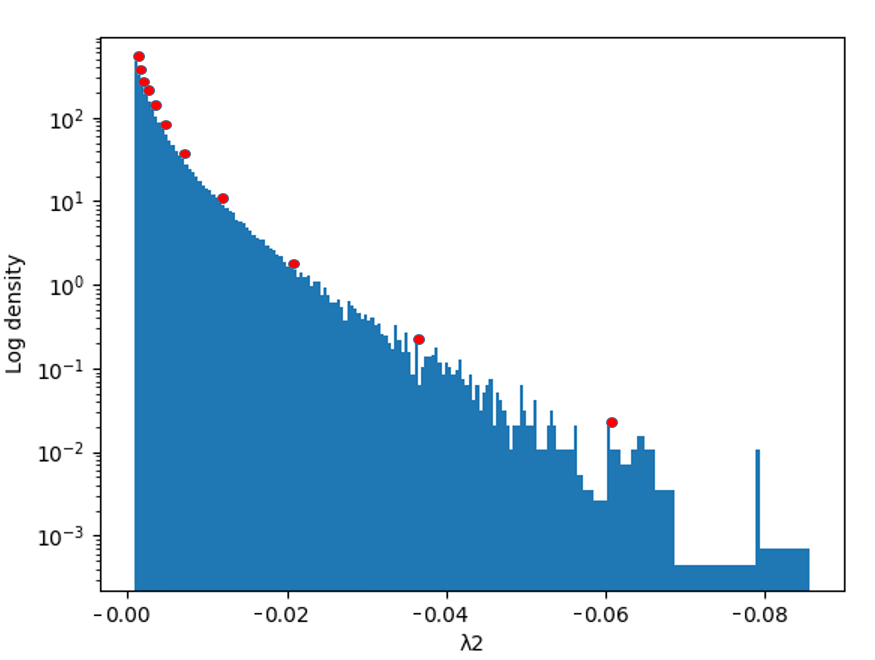}
         \vskip -4pt
         \caption{}
     \end{subfigure}
     \vskip -5pt
     \begin{subfigure}[b]{0.49\linewidth}
         \centering
         \includegraphics[width=0.99\linewidth]
         {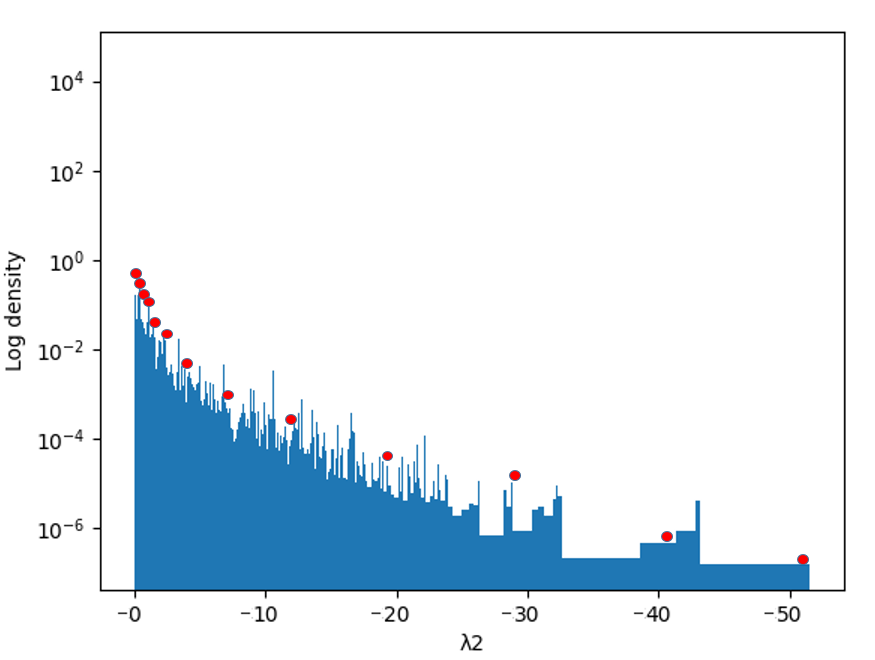}
         \vskip -4pt
         \caption{}
     \end{subfigure}
          \begin{subfigure}[b]{0.49\linewidth}
         \centering
         \includegraphics[width=0.99\linewidth]{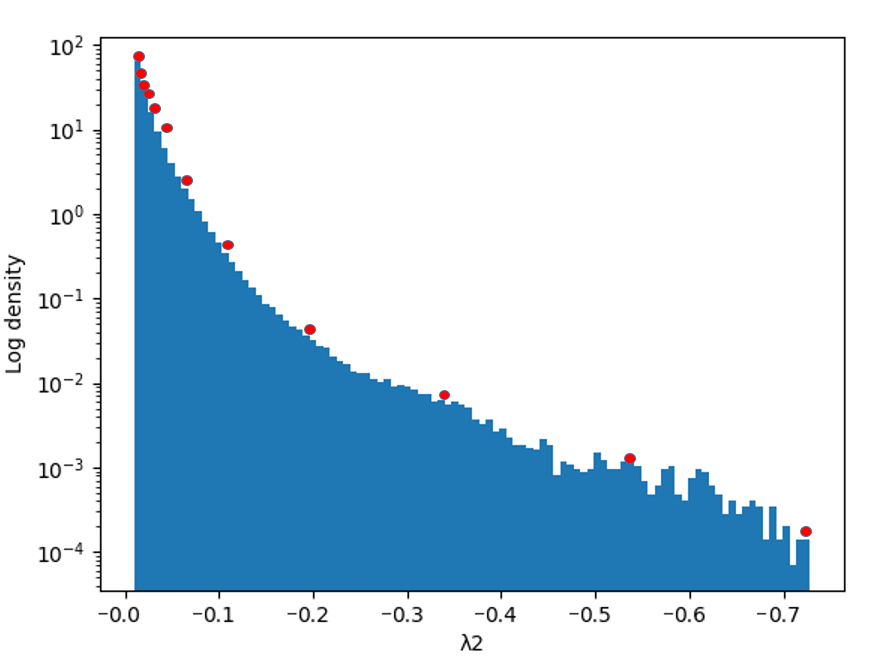}
         \vskip -4pt
         \caption{}
     \end{subfigure}
     \vskip -3pt
 \subfigsCaption{This figure shows the finalized histograms using the histogram expansion approach for (a) B\'enard (b) Crayfish (c) Plume and (d) Couette flow in a log scale. It can be seen that the concentration of values is towards the larger $\lambda_2$. Red dots show the indices of the bins being picked using the Fibonacci series.
 }
 \label{fig:fig3.2.1}
\end{figure}

We use Fibonacci series because the percentage of voxels with larger values of $\lambda_2$ is greater than the lower values as shown in \cref{fig:fig3.2.1}. Therefore more $\lambda_2$ values should be picked from the bins having higher probabilities as compared to the bins with lower probabilities. Any other mathematical function that exhibits this incremental property would work equally fine. \revise{The number of indices in $\lambda_2\_steps$ is equal to the depth of the tree.} One could argue that extracting the complete contour tree with all the critical points and pruning its nodes afterwards can result in equally valid vortices as was done in \cite{bremer2015identifying}. This is valid. However, as mentioned earlier, the complete contour tree is very dense, and the computational overhead of pruning the dense graph is higher as compared to the sparse one. Picking the $\lambda_2\_steps$ using our histogram expansion approach results in a sparse tree and any pruning (if needed) is less computationally expensive simply due to the fact that the resulting tree has fewer nodes.

\begin{figure}[t]
 \centering 
      \begin{subfigure}[b]{1\linewidth}
         \centering
         \includegraphics[width=0.45\linewidth]{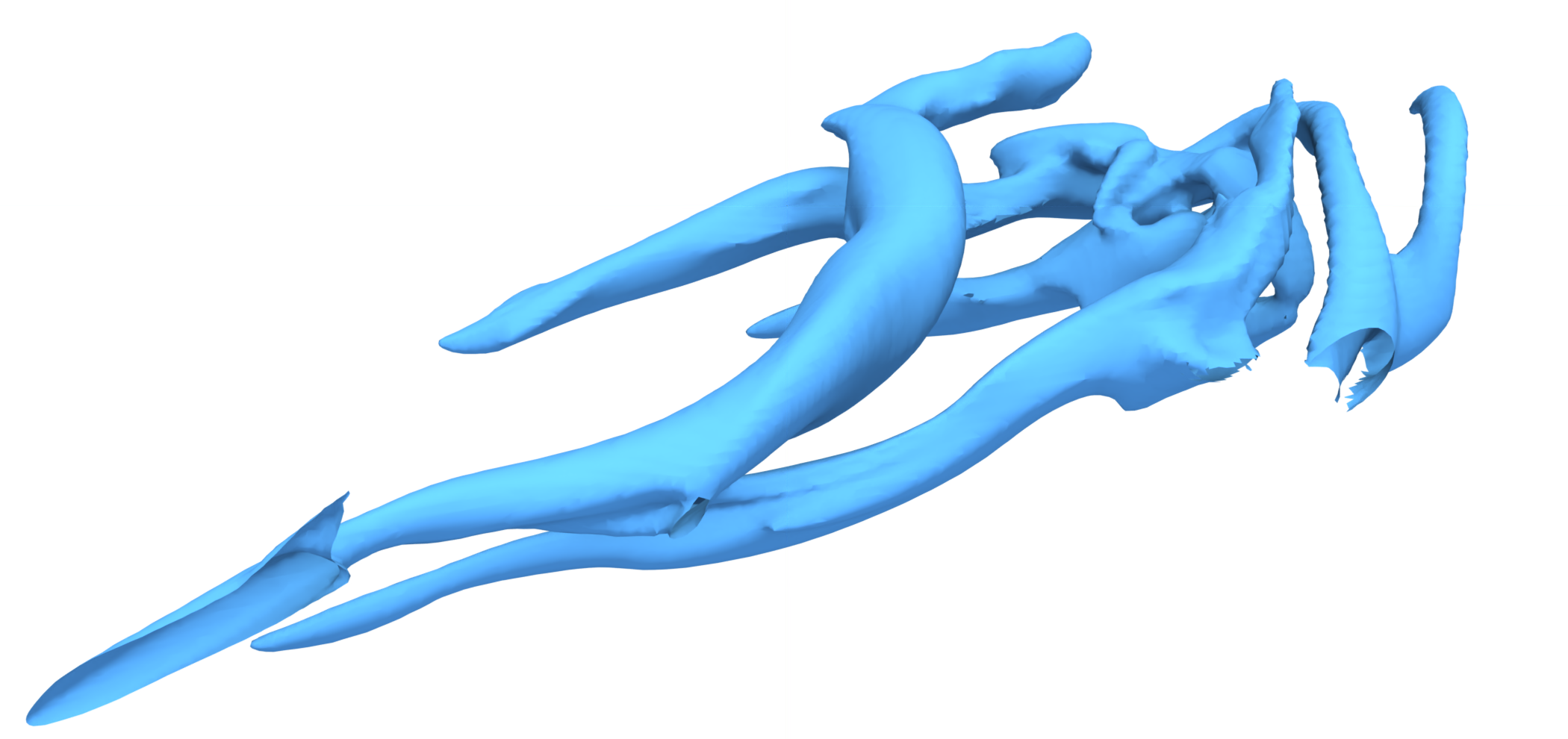}
         \includegraphics[width=0.45\linewidth]{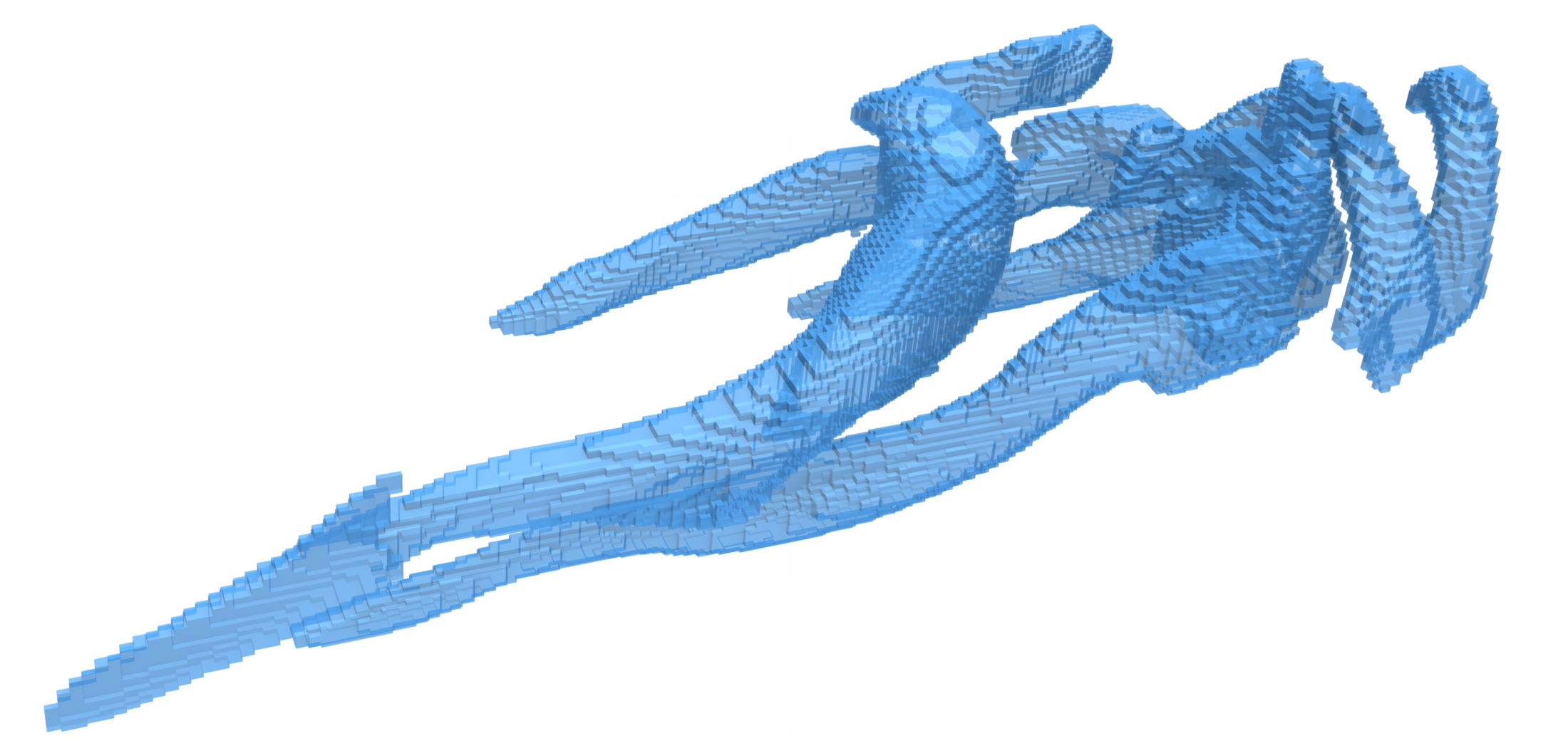}
         \vskip -5pt
         \caption{}
         \label{fig:fig3.2.3a}
     \end{subfigure}
     \vskip -2pt
     \begin{subfigure}[b]{1\linewidth}
         \centering
         \includegraphics[width=0.45\linewidth]{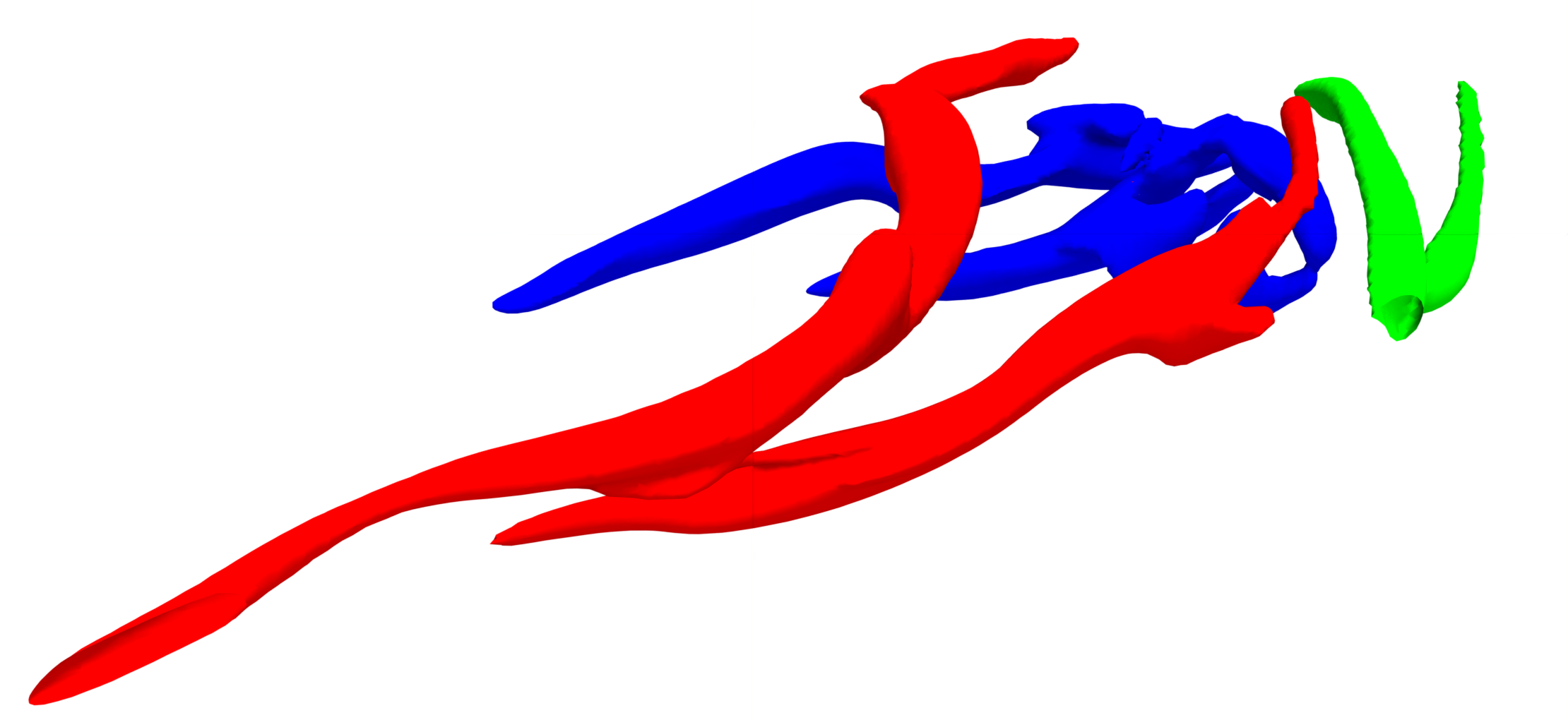}
         \includegraphics[width=0.45\linewidth]{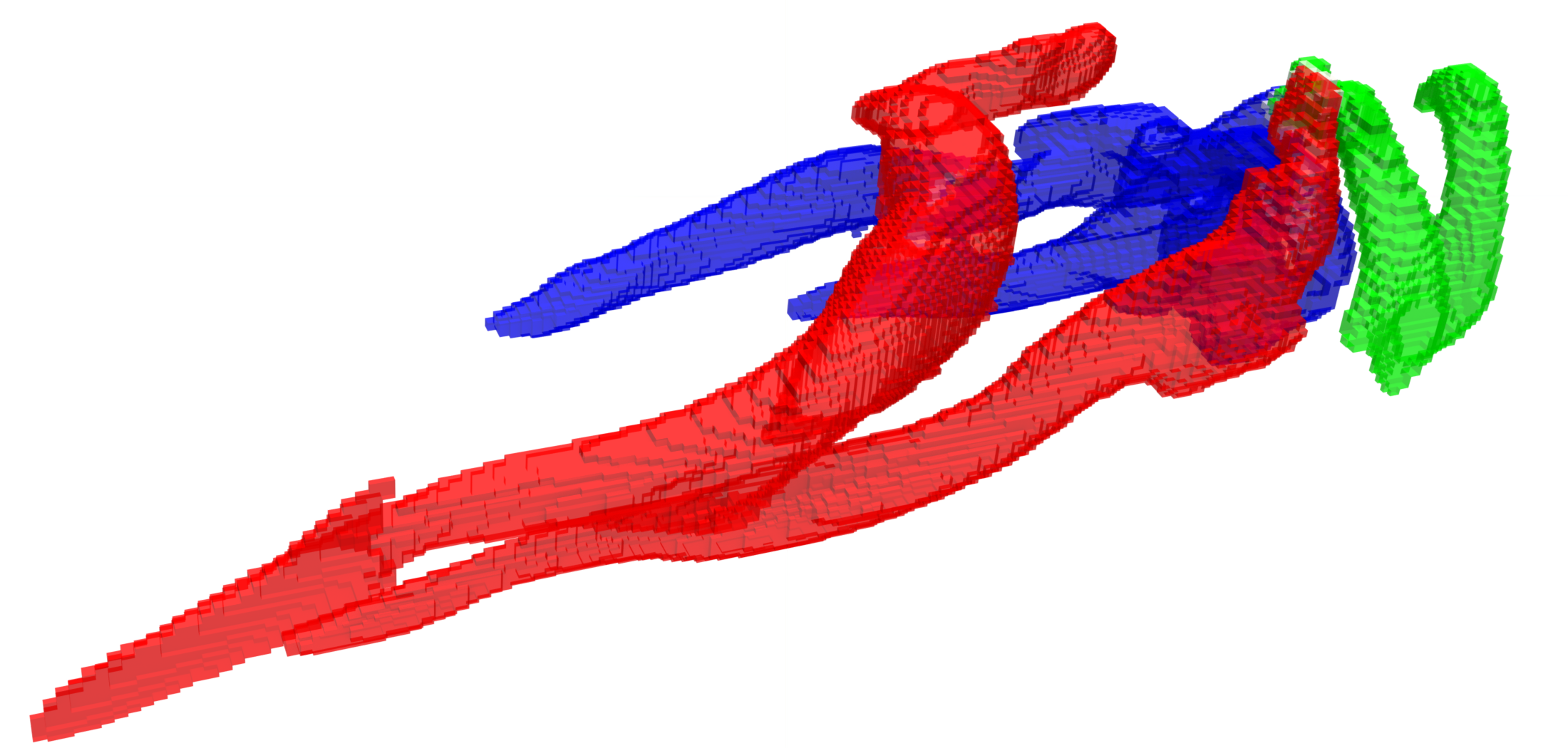}
         \vskip -5pt
         \caption{}
     \end{subfigure}
     \vskip -2pt
 \subfigsCaption{This figure shows the region-splitting process based on the underlying $\lambda_2$ iso-surfaces. The left side shows the extracted isosurfaces and the right side shows the extracted unstructured grid using the region-growing strategy. In (a) there is only one iso-surface component so the region is not split (same color). In (b) there are 3 iso-surface components (different colors) so the region gets split into 3 children. 
 }
 \label{fig:fig3.2.3}
\end{figure}

Given $\lambda_2\_steps$ described above and the vortical regions $\{R\}$ extracted using the region growing strategy mentioned in \cref{section3.1}, we use a top-down region-splitting strategy to split the regions. For each $\lambda_2$ value in $\lambda_2\_steps$, we extract an isosurface and check if the iso-surface has one or multiple parts/components. If the isosurface has only one component, we do nothing and pick the next value from $\lambda_2\_steps$ and extract a new isosurface. If the isosurface has multiple components then we split a region $R$ as follows. We first assign a unique \textit{'color id'} to each component of the isosurface. Then for each individual cell in $R$, we assign it the \textit{'id'} same as \textit{'color id'} of the isosurface component to which it is closest depending on the Euclidean distance. This process is shown in \cref{fig:fig3.2.3}. Generally, if there are $m$ isosurface components, the region will get split into $m$ smaller regions. We repeat this process for each value in $\lambda_2\_steps$. \revise{In region splitting, we ignore the isosurface components which are smaller than a certain number of cells. In our implementation, we set this to 0.01\% of the total number of voxels in the dataset.}

Since each region $R \in \{R\}$ is an independent unstructured grid, the splitting can be performed in parallel. This process results in a hierarchical tree of vortical regions. The tree's level represents a particular $\lambda_2$ isovalue that caused the region to split at that level and nodes represent the vortical regions. The hierarchical relation between the nodes represents the parent regions and the child regions that it got split into. The leaf nodes show the final vortical regions that cannot be split anymore.

\noindent\revise{\textbf{Controlling the Tree's Density}: There are two ways to control the density (or the number of nodes) of the tree.  Firstly, by the choice of parameters in steps (i) and (ii) of the histogram expansion strategy above, increasing the number of bins in step (i) pushes the smaller $\lambda_2$ values toward the higher bins. This decreases the $\lambda_2$ value corresponding to each index in $\lambda_2\_steps$, therefore, resulting in bigger step sizes. Due to the bigger step sizes, the vortices will split less and the tree will be sparser. Conversely, increasing the stop threshold and the factor to increase the number of bins, in step (ii), increases the corresponding $\lambda_2$ values in $\lambda_2\_steps$. This results in smaller step sizes and consequently a denser tree. Secondly, in order to avoid degenerate splits which consequently control the tree's density, we define two ratios.} The ratio between the length of the isosurface component and the length of the region $R$ is called the local length ratio denoted as $L_r$. Here ``length'' means the length of the diagonal of the bounding box of the object. The ratio between the length of the region and the length of the dataset is called the global length ratio denoted as $G_r$. We ignore the isosurface components which have $L_r < \frac{VSF}{G_r}$, here $VSF$ is prolonged as the \textbf{V}ortex \textbf{S}ize \textbf{F}actor. Flow datasets can have small vortices in comparison with the length of the dataset and vice versa. $VSF$ allows the user to control the size of the extracted vortices in the dataset, e.g., in the Plume and Crayfish flows, the vortices are smaller in size when compared with the length of the dataset, thus $VSF$ should be set to a smaller value; and in Cylinder and B\'enard flows, the vortices are larger in size, therefore $VSF$ should be set to a larger value. Moreover, $L_r$ is inversely proportional to $G_r$. At the start of the region splitting process, $G_r$ is relatively large, therefore it allows the isosurface components with smaller $L_r$ to split. When regions get split into smaller regions, $G_r$ decreases and it only allows relatively larger isosurface components to split. This is because, at smaller scales, we want more accurate splits. \revise{It is recommended to use $VSF$ to control the density of the tree. The default value of 3.5 is used in our experiments for all the datasets mentioned in \cref{tab:Table2}. For a general turbulent flow, we recommend starting with the default $VSF$ value and subsequently increasing/decreasing the value to increase/decrease the amount of split if necessary.}

\begin{figure}[h]
 \centering 
 \includegraphics[width=0.9\linewidth]{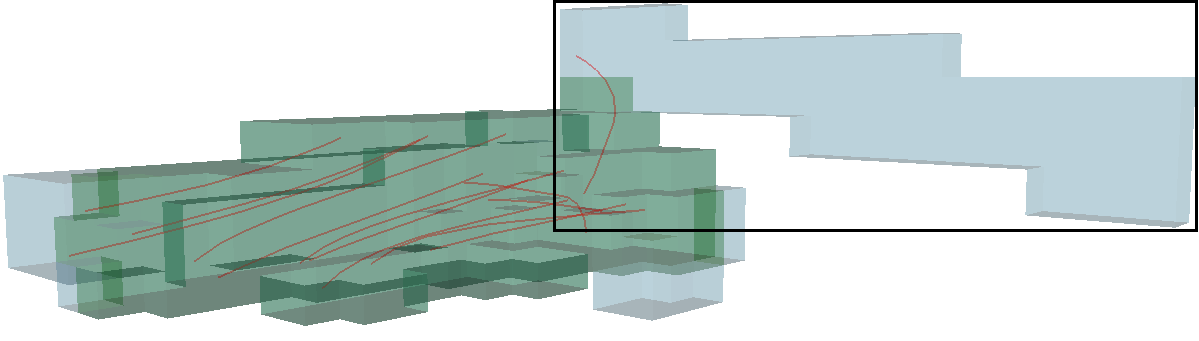}
 \subfigsCaption{This figure shows an extracted vortical region. The red lines are vortex lines. The cells highlighted in black are ill-connected and have very low vorticity. After the region simplification, we are only left with the green region. It can be seen that the simplified region is more homogeneous and have strong vorticity which is depicted by the multiple vortex lines (red).}
 \label{fig:fig3.2.5}
\end{figure}

\noindent\textbf{Region Simplification}: The region growing and splitting approach may result in vortical regions of irregular shapes. This is more prominent in small-scale and coarse-grained datasets, e.g. B\'enard flow. We use geometric and physical criteria to simplify such ill-structured regions into accurate shapes. We remove all the cells having less than 6 neighbors. For all the remaining cells, we remove the ones having $\lambda_2$ value less than the average $\lambda_2$ value of the region and the number of cell neighbors less than 8. The values of 6 and 8 above are user-controlled. \revise{The default values provide a minimal necessary amount of simplification based on our experiments on datasets in \cref{Tab:table1}}. One simple example of region simplification is shown in \cref{fig:fig3.2.5}.

\section{Vortex Profile Construction}
\label{sec:vortexprofiling}

We construct a profile for each separated vortex. The profile includes its geometric features, such as size, length, orientation, curvature, etc. The vortex profile also contains important physical properties of the vortex such as the average vorticity, enstrophy, and others (such as $Q$, $\lambda_2$, etc.). The list of all the features included in the vortex profile is shown in \cref{Tab:table1}. This profile forms a feature vector that statistically represents this vortex. With this representation, 
(1) we can calculate the statistics of the profiles of all the vortices. For example, we can find the vortex having the average, minimum (and/or maximum) values of certain features. (2) We can sort and visualize the subset of vortices by specifying the range of a certain feature in the vortex profile. Furthermore, (3) we can project each vortex to a low-dimensional space based on their respective signature vectors using a certain dimensionality reduction technique. We then can perform clustering to classify vortices into different clusters depending on their features.

\begin{figure}[t]
 \centering
      \begin{subfigure}[b]{0.3\linewidth}
         \centering
         \includegraphics[width=0.99\linewidth]{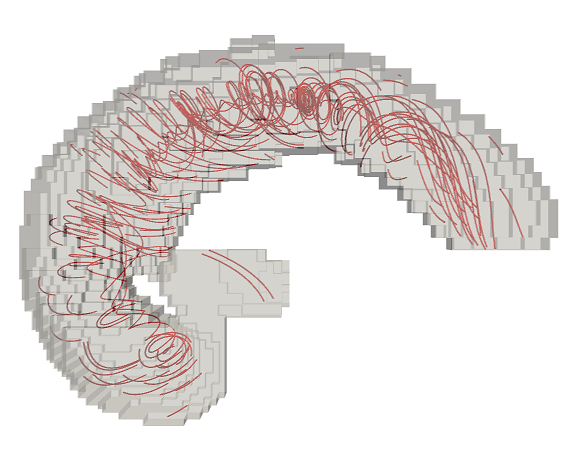}
         \vskip -5pt
         \caption{}
     \end{subfigure}
    \begin{subfigure}[b]{0.3\linewidth}
         \centering
         \includegraphics[width=0.99\linewidth]{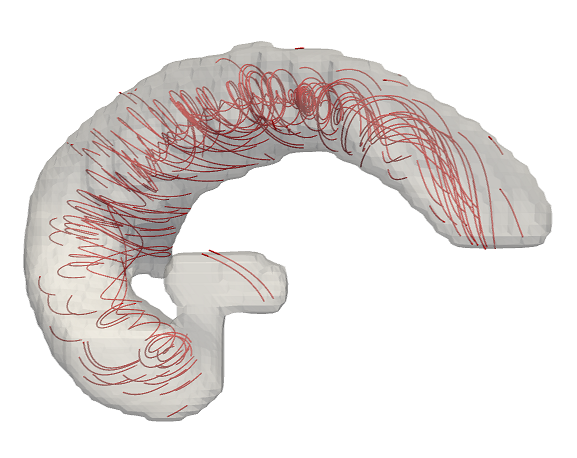}
         \vskip -5pt
         \caption{}
     \end{subfigure}
    \begin{subfigure}[b]{0.3\linewidth}
         \centering
         \includegraphics[width=0.99\linewidth]{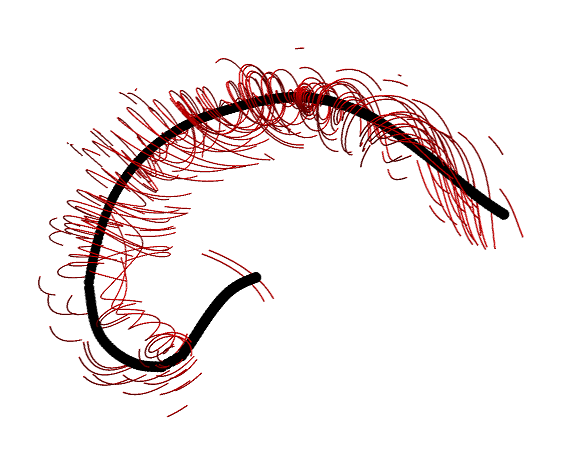}
         \vskip -5pt
         \caption{}
     \end{subfigure}
     \vskip -2pt
 \subfigsCaption{This figure shows the skeletonization process of an extracted vortical region. (a) shows the grid of the vortical region, (b) shows the smooth surface of the grid, and (c) shows the extracted skeleton (black). Streamlines (red) verify that the grid is a vortical region.}
 \label{fig:fig4.1}
\end{figure}

These functionalities are  further demonstrated in \cref{sec:visualizationsystem}. Since the extracted vortices are unstructured grids, approximating geometric features (such as shape, orientation etc.) is not straightforward. Therefore, we resort to skeleton extraction to approximate such geometric features. We extract vortex skeletons using the mean curvature skeletons (MCFSKEL) \cite{taglia_sgp12}. We first extract the outer surface of the vortical region, smooth it and apply MCFSKEL to extract the skeleton. We also tried using vortex corelines for shape approximation as done in \cite{adeel2022hairpin}, but vortex corelines might be broken and disconnected. We use the surface of the unstructured grid instead of $\lambda_2$ isosurface because the $\lambda_2$ isosurface may not cover the entire region due to the isovalue used to split that region. The $\lambda_2$ isosurface may also have boundaries. The surface of the unstructured grid is complete and closed. Since the outer surface of an unstructured grid can be bumpy depending on the cell type, therefore we perform surface smoothing. \revise{We use 10 iterations of Laplacian Smoothing \cite{field1988laplacian} to keep the underlying shape intact and reduce the bumps minimally. With less than 10 iterations, the bumps are significantly visible, and with more than 20 iterations, the mesh starts to shrink based on our experiments.} The skeletonization process is shown in \cref{fig:fig4.1}. 

\begin{table}[!b]
\caption{This table shows the features used for constructing vortex profiles. \revise{The geometric features (last 4 rows) are normalized with the number of points in the skeleton. The physical features (first 6 rows) are the volume averages of the vortical (unstructured) grids}. For vector and tensor quantities such as velocity and Jacobian, we take the average of the magnitude and norm respectively. This results in a 19-dimensional feature vector for each vortex\textsuperscript{\ref{xx}}.}
\scriptsize%
\centering
\vspace{0.01in}
\begin{tabular}{ |p{0.3\linewidth}|p{0.1\linewidth}||p{0.3\linewidth}|p{0.1\linewidth}| }
 \hline
 \multicolumn{4}{|c|}{Vortex Profile Features List} \\
 \hline
Name & Symbol & Name & Symbol\\
 \hline
 $Lambda_2$ & $\lambda_2$ & $Lambda_{ci}$ & $\lambda_{ci}$ \\
 Q & $Q$ & Delta & $\Delta$ \\
 Divergence & $Div$ & Spanwise Vorticity Fluctuations (Oyf) &  $\omega_{y}'$ \\
 Size \revise{(\# of Voxels)} & $Size$ & Vorticity &  $\omega$ \\
 Enstrophy & $\xi$ & Velocity & $V$ \\
 Acceleration & $a$ & Jacobian & $J$ \\
 Curvature & $C$ & Hairpin Curvature (eq.\ref{eq3}) & $\tilde{C}_h$\\
 Streamwise Direction & $S_t$ & Spanwise Direction & $S_p$\\
 Vertical Direction & $S_v$ & Length (eq.\ref{eq3}) & $L$\\
 Bbox Ratio (eq.\ref{eq3}) & $\rho$ & - & -\\
 \hline
\end{tabular}
\label{Tab:table1}
\end{table}

Last four rows of \cref{Tab:table1} represent the shape and orientation information of the skeleton. Curvature ($C$) is calculated as the normalized sum of angles between the normalized tangents on adjacent pairs of points on the skeleton. \revise{The variable \emph{Streamwise Direction} ($S_t$) quantifies the vortex direction towards streamwise}. It is calculated as the normalized sum of dot products between the unit vector in streamwise direction and the normalized vectors formed by the line segments between two adjacent points in the skeleton. Similarly, \emph{Spanwise Direction} ($S_p$) is computed with the unit vector in the spanwise direction and \emph{Vertical Direction} ($S_v$) with the unit vector of the 3rd direction which is perpendicular to both streamwise and spanwise directions.

After constructing the profiles for individual vortices, all or subsets of the attributes in the profiles can be used to characterize the different types of vortices. Next, we concentrate on hairpin vortices and their characterization.
\section{Hairpin Vortex Identification}
\label{sec:hairpin}

In order to accurately identify hairpin vortices, several factors need to be accounted for. Generally, at low Reynolds numbers, hairpin vortices have relatively well-defined shapes and separate from each other. At medium and high Reynolds numbers, hairpin vortices start to take different forms due to the viscous stress imposed by other close-by coherent structures \cite{adrian2007hairpin}. In the literature, hairpin vortices are sometimes also called "Arches", "Horseshoes" or "Omega-shaped" vortices due to the fact that the hairpin may not have legs and/or have a single leg\cite{adrian2007hairpin}. Moreover, the organization of the hairpin is not a single-step phenomenon. During the whole lifespan of a hairpin vortex, it goes through different shapes and forms. Three major stages in the hairpin vortex lifespan are: (a) Formation, (b) Roll Up and (c) Breakdown, details in \cite{kim1987turbulence}\textsuperscript{\ref{xx}}. During stages (a) and (b), the spanwise vorticity fluctuations $\omega_{y}'$ have a strong positive value in the head of the hairpin vortex \cite{adrian2007hairpin}. Therefore, we make $\omega_{y}'$ as the primary physical feature to detect hairpin vortices. In stage (a), the hairpin vortex has an "arch" and at stage (b), the shape is like a "Hairpin" if both legs are present or an "Omega" like shape if the legs are partial or not present entirely. If there is only one leg, the head still has an "arch". To quantify "arch", we calculate the curvature $C$ of the skeleton of the vortex in \cref{Tab:table1} and use it as the primary geometric feature. We also use $S_t$, $S_p$ and $S_v$ in \cref{Tab:table1} as the secondary geometric features for the task.

\revise{In the Couette flow dataset,} the wall with the no-slip condition\footnote{\label{xx}Readers can refer to the supplemental document for additional details.} is better suited for the generation of hairpin vortices \cite{li2019direct}. Therefore, in order to identify hairpin vortices, we first put more emphasis on the vortices close to the no-slip \emph{bottom boundary} in the Couette flow \cite{li2019direct}. The steps to identify hairpin vortices are given below.

\begin{figure}[t]
 \centering
    \begin{subfigure}[b]{0.49\linewidth}
         \centering
         \includegraphics[trim={0 5cm 0 5cm},clip,width=0.99\linewidth]{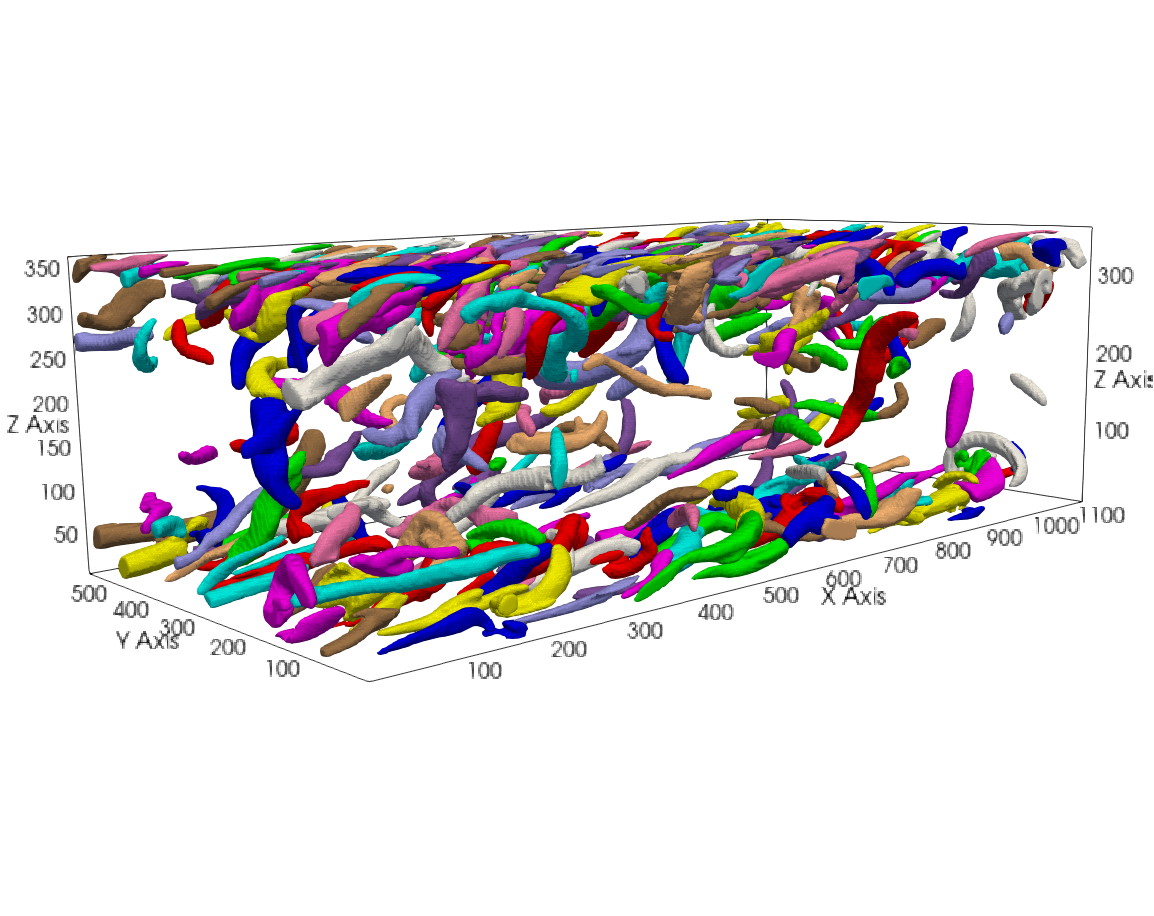}
         \vskip -5pt
         \caption{}
         \label{fig:fig6.1a}
    \end{subfigure}
    \hfill
    \begin{subfigure}[b]{0.49\linewidth}
         \centering
         \includegraphics[trim={0 5cm 0 5cm},clip,width=0.99\linewidth]{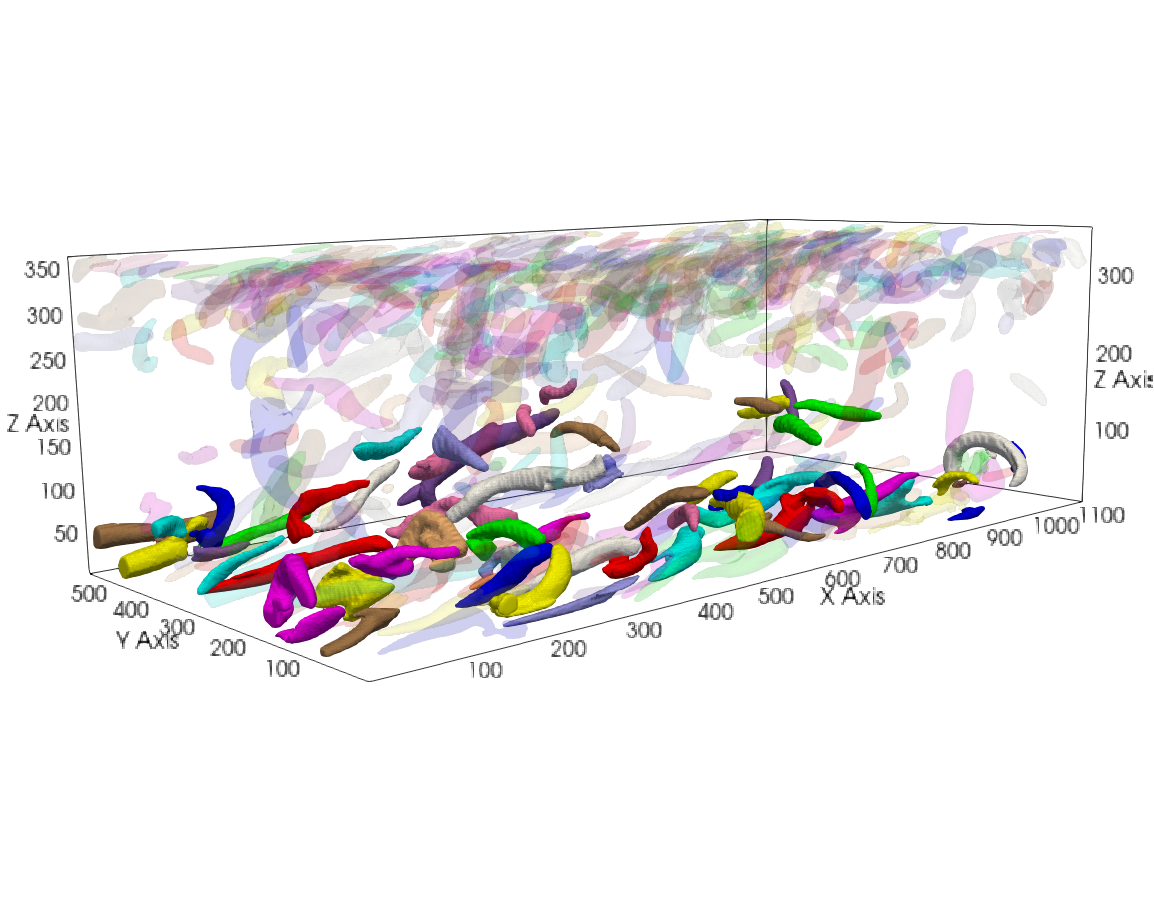}
         \vskip -5pt
         \caption{}
         \label{fig:fig6.1b}
     \end{subfigure}
     \hfill
     \vskip -5pt
     \begin{subfigure}[b]{0.49\linewidth}
         \centering
         \includegraphics[trim={0 5cm 0 5cm},clip,width=0.99\linewidth]{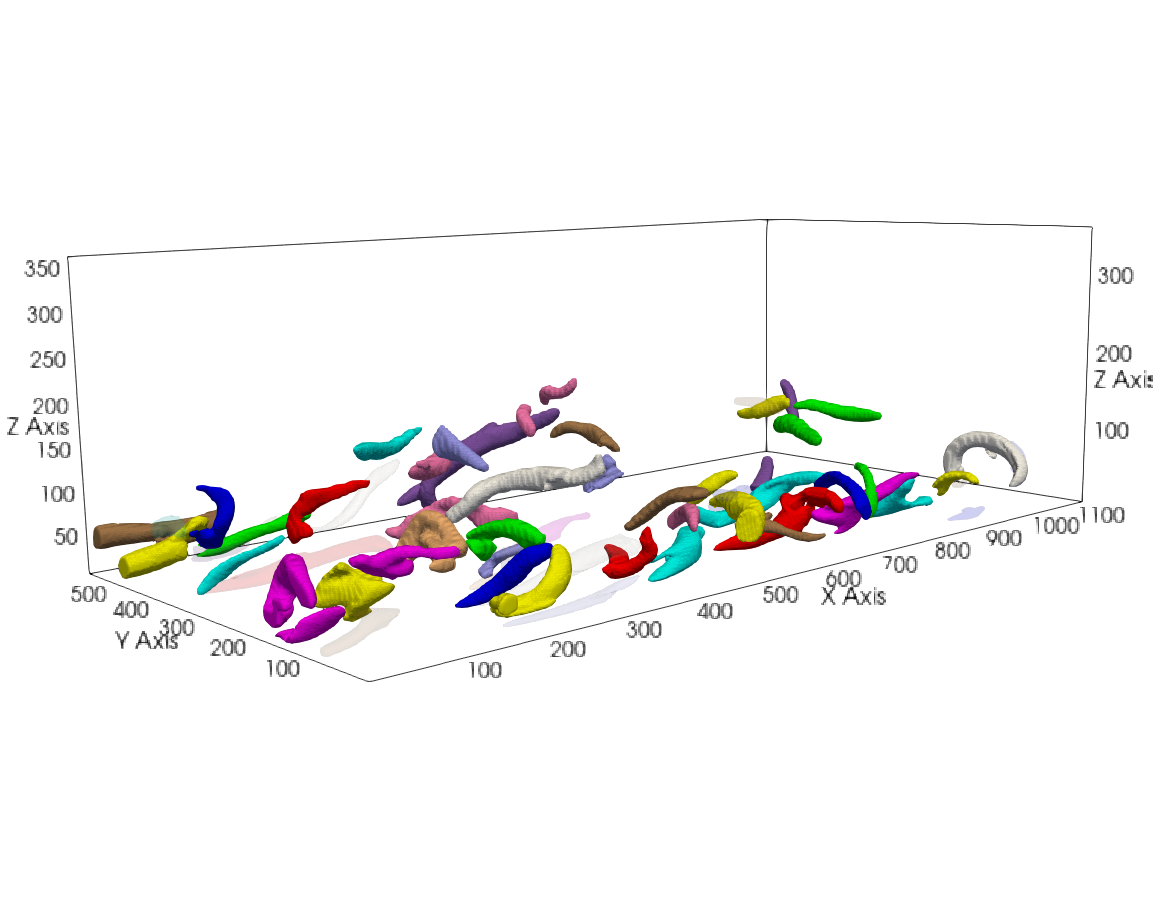}
         \vskip -5pt
         \caption{}
         \label{fig:fig6.1c}
     \end{subfigure}
          \begin{subfigure}[b]{0.49\linewidth}
         \centering
         \includegraphics[trim={0 0cm 0 0cm},clip,width=0.99\linewidth]{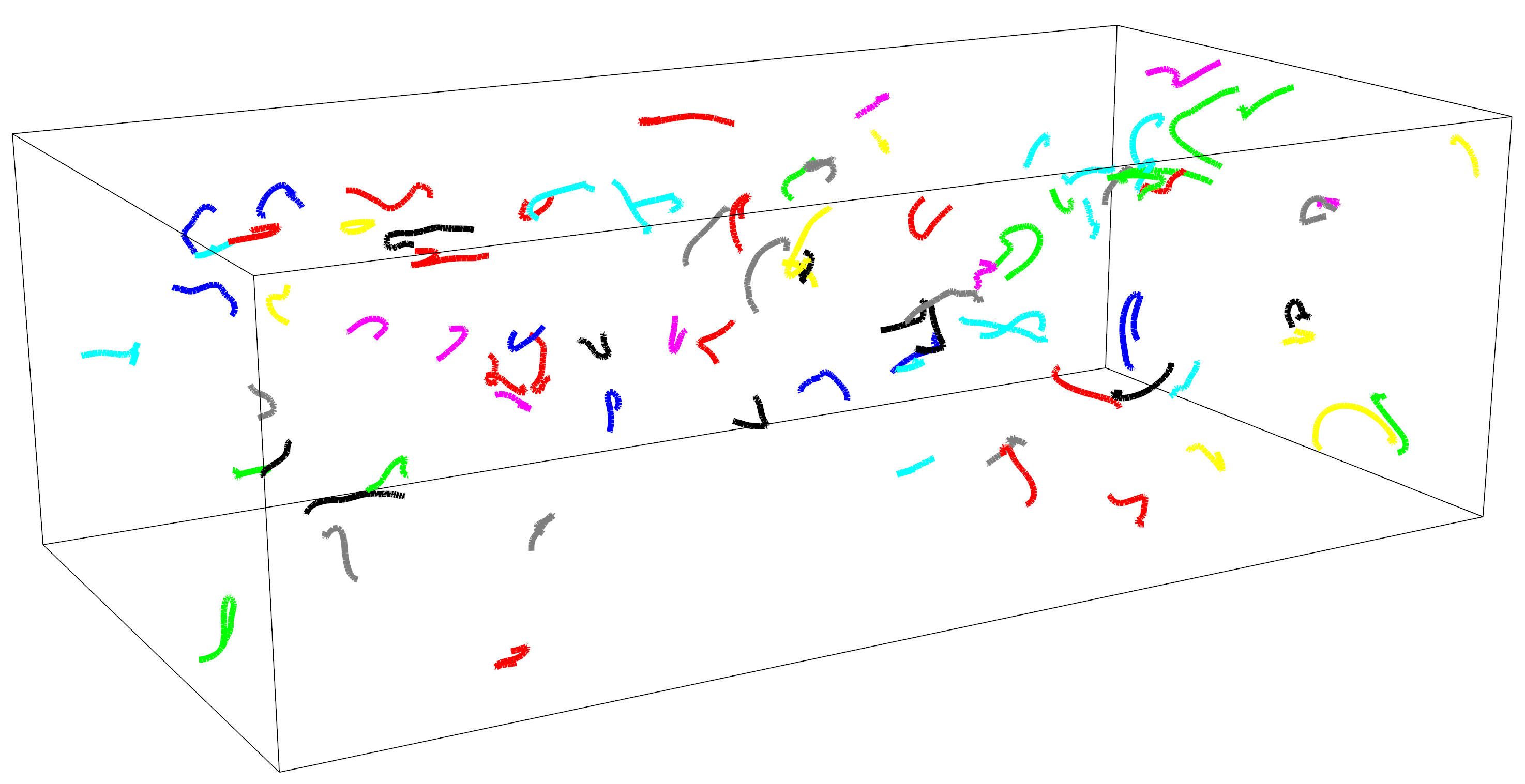}
         \vskip -5pt
         \caption{}
         \label{fig:fig6.1d}
     \end{subfigure}
     \vskip -2pt
 \subfigsCaption{This figure shows the effect of applying Steps 1, 2, and 3 to identify candidate hairpin vortices in the Couette flow. (a) shows all the vortices, (b) shows the effect of applying Steps 1 and 2, and (c) shows the candidate hairpin vortices. Blurring in (b) and (c) shows that those vortices have been filtered out. Random colors are assigned to differentiate between vortices. (d) shows the results of the work \cite{adeel2022hairpin} for comparison purposes. (Best viewed in zoom).
 }
 \label{fig:fig6.1}
\end{figure}

\begin{figure*}[t]
 \centering 
    \begin{subfigure}[b]{0.48\textwidth}
         \centering
         \includegraphics[width=0.95\textwidth]{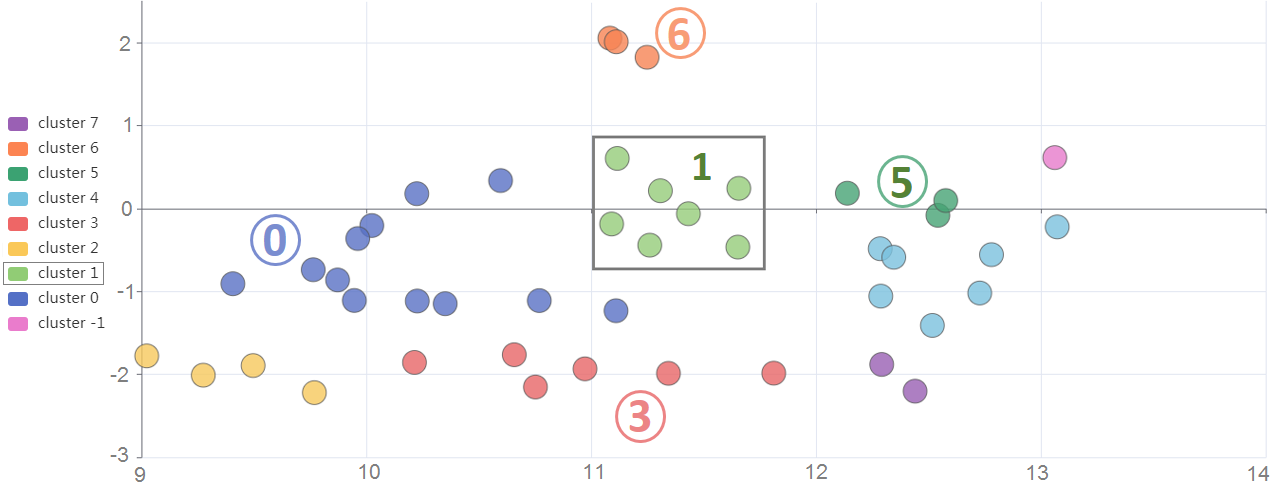}
         \vskip -2pt
         \caption{}
         \label{fig:fig5.1a}
     \end{subfigure}
     \hfill
    \begin{subfigure}[b]{0.48\textwidth}
         \centering
         \includegraphics[width=0.95\textwidth]{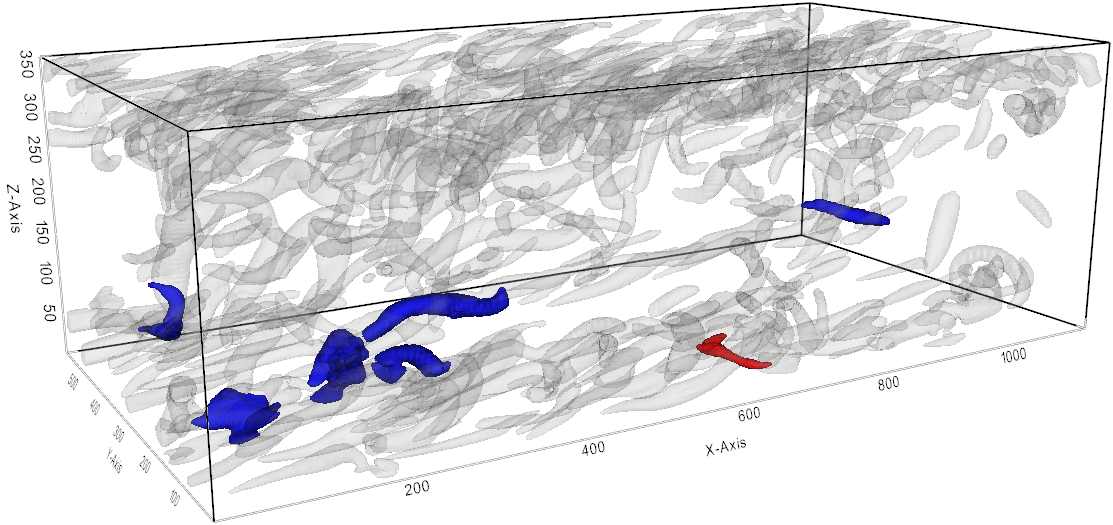}
         \vskip -2pt
         \caption{}
         \label{fig:fig5.1b}
    \end{subfigure}
     \vskip -3pt
     \begin{subfigure}[b]{0.48\textwidth}
         \centering
         \includegraphics[width=0.95\textwidth]{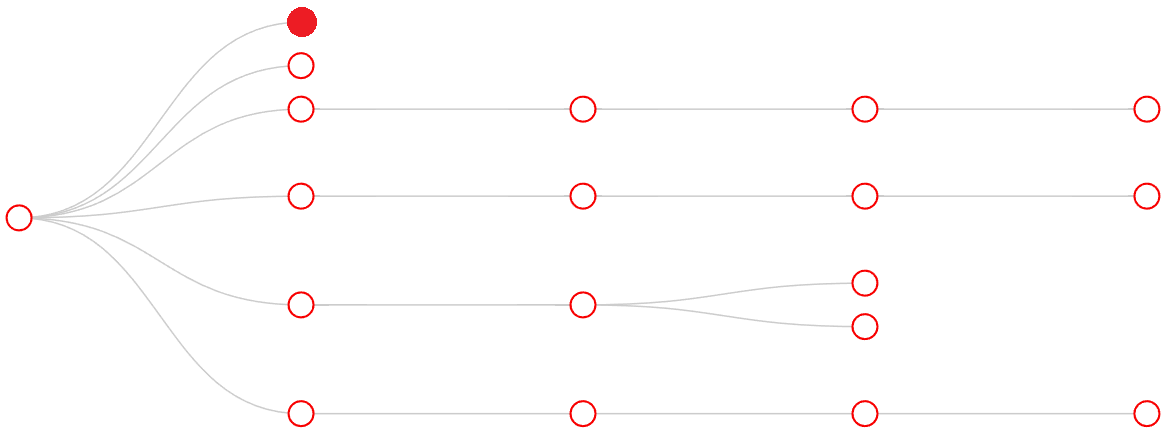}
         \vskip -2pt
         \caption{}
         \label{fig:fig5.1c}
     \end{subfigure}
     \hfill
     \begin{subfigure}[b]{0.48\textwidth}
         \centering
         \includegraphics[width=0.95\textwidth]{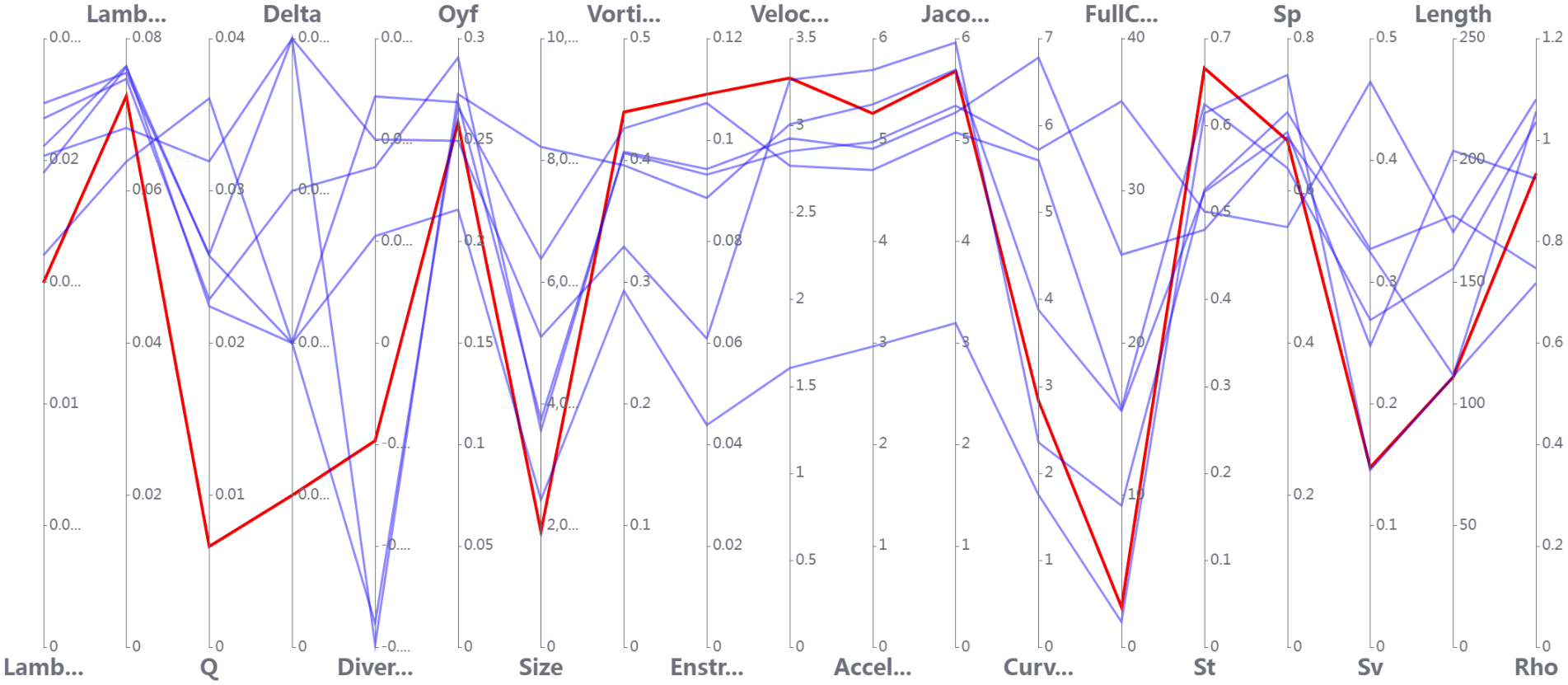}
         \vskip -2pt
         \caption{}
         \label{fig:fig5.1d}
     \end{subfigure}
 \subfigsCaption{This figure shows an example interaction between (a) Cluster View, (b) 3D View, (c) Tree View, and (d) PCP View. The vortices highlighted in 3D View (b), the lines in PCP view (d), and the leaf nodes in the Tree View (c) correspond to cluster 1 highlighted in (a). The node highlighted in red is selected by the user in the tree view (c), and the corresponding vortex and profile is highlighted in the 3D view (red) and the PCP view (red line). The tree (c) shows the hierarchy of the vortices in the selected cluster. We highlight some clusters with their respective \emph{id} for reference purposes.
 }
 \label{fig:fig5.1}
\end{figure*}

\vspace{0.05in}
\noindent \textbf{Step 1: } We discard all the vortices with $S_t > T_1$, $S_p > T_1$, and $S_v > T_1$, meaning filter out all the vortices which are entirely in the streamwise, spanwise or vertical direction. Again by vertical, we mean the direction perpendicular to the streamwise and spanwise. Where $T_1$ is a user-specified parameter which we set to 0.99 in our implementation. 

\vspace{0.05in}
\noindent \textbf{Step 2: } We remove vortices based on their spanwise vorticity fluctuations ($\omega_y'$). Since we focus on the vortices near the bottom boundary, to suppress the vortices that have high $\omega_y'$ and are far away from the bottom, we first multiply the original $\omega_y'$ by $t=1-\frac{z-Z_{min}}{Z_{max}-Z_{min}}$ ($Z_{min}$ and $Z_{max}$ are the z coordinates for the bottom and top boundaries of the domain, respectively). 

Note that $\omega_{y}'$ is calculated at the individual points and can be both positive and negative $\omega_{y}'$. If we directly calculate the average based on all points within a vortex, the negative values decrease the overall average. To avoid this problem, we calculate the root mean square (RMS) of $\omega_{y}'$ along the vortex skeleton. There are a few considerations while calculating the RMS. (a) If all the points of the vortex skeleton have a negative $\omega_{y}'$, the RMS will still be high. (b) One can calculate the average of the points with positive $\omega_{y}'$ only; but if only a small percentage of the points have positive $\omega_{y}'$, their average will be high and we completely ignore the effect of the points with negative $\omega_{y}'$. (c) On the other hand, if we consider only the positive $\omega_{y}'$ values and divide them by the total number of points, the RMS will get smaller even for the vortices having strong sections of high $\omega_{y}'$. To address these issues, we calculate the RMS using equation \cref{eq1}.
    \begin{equation}
        \label{eq1}
        RMS(\omega_{y}') = \sqrt{\frac{1}{n}\sum(\omega_{y}'')^2},
        \quad \omega_{y}'' = 
        \biggl\{
            \begin{array}{lr}
                \omega_{y}', & \textit{if} \; \omega_{y}' \geq 0 \\
                0.1 \times \omega_{y}', & \textit{if} \; \omega_{y}'< 0
            \end{array}
    \end{equation}   
    where $n$ is the number of points in the skeleton. This will reduce the effect of negative $\omega_{y}'$ values while retaining the importance of positive $\omega_{y}'$ values. We filter out the vortices with $RMS(\omega_{y}') \leq 0$.

\noindent \textbf{Step 3: }We remove vortices having low curvature (i.e., those flat vortices).  To achieve that, we follow the properties of the hairpin vortices from \cite{kim1987turbulence} and \cite{adrian2007hairpin}. At stage (a), the direction of the hairpin vortices exhibits a high spanwise component $S_p$. During the roll-up stage, hairpin vortices thrust in a direction perpendicular to the wall. Therefore, they also exhibit a high vertical direction component $S_v$. Hairpin vortices are less likely to have the major part of their direction purely streamwise since their heads mostly point in the spanwise direction which reduces the overall streamwise component $S_t$. If a vortex has a very high percentage of streamwise direction component $S_t$, it is more likely to be a quasi-streamwise vortex. Therefore, we penalize the vortex curvature having a high $S_t$. We incorporate these direction components to calculate the hairpin curvature $C_h$ using \cref{eq2}.
    \begin{equation}
    \label{eq2}
        C_h = C \times (1-S_t) \times S_p \times S_v
    \end{equation}
    where $C$ is the curvature calculated in \cref{Tab:table1}. Next, we calculate the oriented bounding box surrounding the vortex skeleton, denoted as \emph{$bbox$}, and the length of the skeleton, $L$, which is calculated as the sum of the Euclidean distances between the adjacent points. We then calculate $\rho=\frac{L}{||bbox||}$ ($||bbox||$ is the length of the diagonal of the $bbox$). $\rho$ is close to 1 for straight skeletons, while larger than 1 for the skeleton having a larger curvature. This is another way to quantify high curvature. We reward a vortex for having a large $\rho$. Moreover, we also incentivize a vortex for having a longer skeleton as given in \cref{eq3}.
       \begin{equation}
    \label{eq3}
       \tilde{C}_h = C_h \times \rho \times \frac{L}{\sqrt[4]{n}}
    \end{equation}
where $n$ in the number of points in the skeleton. $\tilde{C}_h$ is a user-specified parameter. To retain as many candidate hairpin vortices as possible, we set $\tilde{C}_h = 1$ in our implementation, which is considerably low as compared to the complete hairpin vortices which we show in \cref{sec:results}. 
In addition, we ignore skeletons having length $L$ less than 1\% of the dimension of the flow domain. \cref{fig:fig6.1} demonstrate the effect of steps 1, 2 and 3 above.

\section{An Interactive Visualization System}
\label{sec:visualizationsystem}
\revise{We develop an interactive visualization system for the domain experts to perform analysis of vortices in fluid flows.} The visualization system is decoupled from the vortex extraction, separation, and profiling of the pipeline, therefore it can be used for the analysis of the hairpin as well as general vortices. \revise{The system provides multiple linked-views, each representing a different perspective of the separated vortices. Each view along with their reason for inclusion is explained below.}

\begin{enumerate}
  \setlength{\itemsep}{0pt}
  \setlength{\parskip}{0pt}
    \item \textbf{3D View:} This view allows the user to select and visualize the vortices directly from the 3D space. \revise{This view provides a traditional way of interacting with objects in 3D space (\cref{fig:fig5.1a}), which is necessary for the majority of visualization systems.}
    
    
    \item \textbf{Parallel Coordinate Plot (PCP) View:} \revise{The PCP view is included to both visualize high dimensional vortex profiles (\cref{fig:fig5.1d}) and compare multiple vortices at the same time. It also helps shortlist the attributes for clustering (\cref{sec:results}).} The PCP view is linked with the 3D volume view. If the user clicks on a particular line/profile in the PCP view, the vortex corresponding to that line gets highlighted in the volume view.
    
    \item \textbf{Tree View:} Tree view is the most important component for user-interaction (\cref{fig:fig5.1c}). \revise{This view is included to enable exploration of the hierarchy of vortices formed during the separation process as mentioned in \cref{sec:vortexextraction}.} Since the number of nodes can be very large depending on the number of extracted vortices, visualizing the whole tree in a limited screen space can result in cluttering. Therefore we resort to extracting and showing the sub-tree instead. The number of nodes for the sub-tree, the criterion to use for sorting the nodes, and the minimum size (i.e., \# of voxels) of the vortex are user-specified. On the sub-tree extraction, the PCP view also gets updated showing the statistics of all vortices corresponding to the current tree nodes. The tree view is linked with the 3D view as well. Whenever the user selects a particular node in the tree, the vortex corresponding to that particular node gets highlighted in the volume.
    
    \item \textbf{Scatter Plot View}: \revise{Since clustering is the final step to identify hairpin vortices, this view is included to visualize the clustering results using a 2D scatter plot (\cref{fig:fig5.1a}). Our system allows the user to select a subset of attributes from the vortex profile to use for clustering.} We adopt the widely used combination of t-SNE\cite{van2008visualizing} and DBSCAN\cite{ester1996density} for dimensionality reduction and clustering of the vortices respectively. Upon selection of a particular cluster from the scatter plot, the PCP plot and the corresponding vortices will get highlighted in the 3D view.
\end{enumerate}

\Cref{fig:fig5.1} demonstrates an example interaction between different views for the analysis of vortices.
\section{Results and Applications}
\label{sec:results}

We have applied our vortex extraction and classification and the developed interactive visualization system to a number of \revise{flow datasets} to assess its effectiveness. \Cref{tab:Table2} provides the statistics of these data sets and their resulting vortices. The parameters used for each data set are also provided. 

\begin{table*}[t]
\caption{This table shows the parameters and performance statistics of the flow datasets used in our experiments. \emph{Time} shows the performance of the region growing and splitting. \emph{PP Time} is the performance measure of the post-processing (PP) steps including the region simplification, surface extraction, skeletonization, and vortex profiling combined (with a serial implementation). The performance is measured on a PC with an Intel Xeon(R) CPU E5-2630 and 32G RAM. \revise{Readers can refer to the supplemental document for the details of the datasets.}
}
\centering
\scriptsize
\begin{tabular}{|p{0.10\textwidth}|p{0.10\textwidth}|p{0.10\textwidth}|p{0.10\textwidth}|p{0.10\textwidth}|p{0.10\textwidth}|p{0.10\textwidth}|p{0.09\textwidth}| }
 \hline
 \multicolumn{8}{|c|}{Parameters Used for each Data Set and Respective Results} \\
 \hline
  \multicolumn{1}{|c|}{Data Set} & \multicolumn{1}{c|}{Size} & \multicolumn{1}{c|}{Initial $\lambda_2$ value($\lambda_{2i}$)} & \multicolumn{1}{c|}{\# of Nodes} & \multicolumn{1}{c|}{\# of Leaf Nodes} & \multicolumn{1}{c|}{Tree Depth} & \multicolumn{1}{c|}{Time (s)} & \multicolumn{1}{c|}{PP Time (s)}\\
 \hline
B\'enard & 248031 & -6.3574 & 66 & 47 & 7 & 16 & 22\\
Cylinder & 565551 & -0.0036 & 120 & 83 & 16 & 23 & 60\\
Crayfish & 6098358 & -0.0012 & 927 & 626 & 9 & 326 & 369\\
Plume & 7984375 & -4.9261 & 363 & 290 & 24 & 1141 & 470\\
Couette & 28164288 & -0.0168 & 532 & 430 & 7 & 430 & 651\\
 \hline
\end{tabular}
\label{tab:Table2}
\end{table*}

\subsection{Application to Turbulent Couette Flow}

We apply our system to the vector field dataset of the numerical simulation of stress-driven turbulent Couette flow by Yang et. al.~\cite{li2019direct}. The dataset is the result of turbulent flows over progressive surface waves. It has a dimension of $384\times384\times193$. We first apply our region growing and splitting strategy to extract vortices from the input flow, which results in $430$ separable vortices. We then constructed the profiles using all the attributes mentioned in \cref{Tab:table1}. We then shortlist the candidate hairpin vortices close to the bottom no-slip boundary using the steps mentioned in \cref{sec:hairpin} as demonstrated in \cref{fig:fig6.1}. We compare the identified candidates \cref{fig:fig6.1c} with those (\cref{fig:fig6.1d}) by a recent hairpin vortex extraction technique based on a shape matching of vortex corelines given a hairpin template \cite{adeel2022hairpin}. We see that our method identifies candidates that include most hairpins with fewer false positives.

In order to further narrow down our search for the hairpin vortices, we use our interactive visualization system to perform clustering using t-SNE and DBSCAN. The clustering result is shown in \cref{fig:fig5.1a}. To represent a good combination of both physical and geometric attributes, we use $\lambda_2$, $\omega_y'$, $\tilde{C}_h$, $S_t$, $S_p$, $S_v$ and $L$ for clustering. Below we showcase several interesting findings corresponding to the highlighted clusters in \cref{fig:fig5.1a}.

\begin{figure}[!t]
 \centering 
      \begin{subfigure}[b]{0.99\linewidth}
         \centering
         \includegraphics[width=0.95\linewidth]{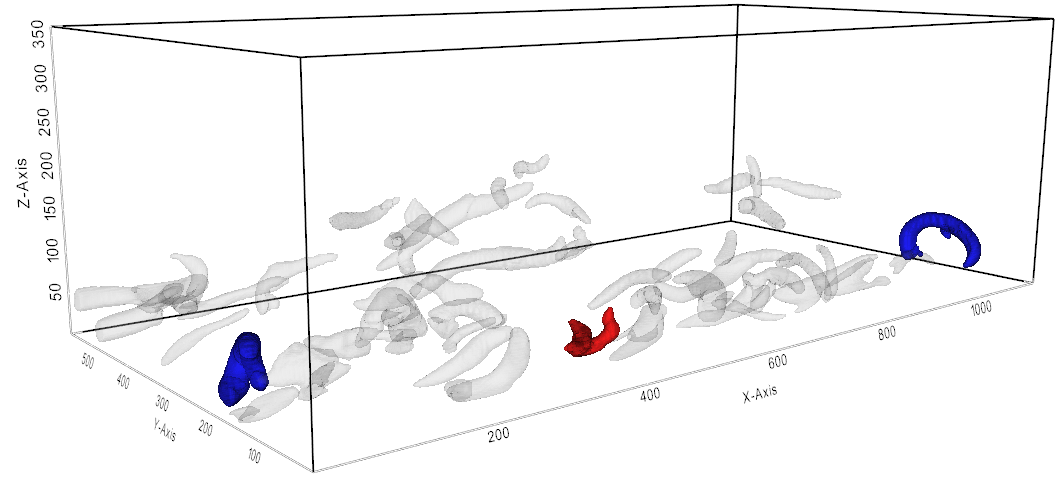}
         \vskip -5pt
         \caption{}
         \label{fig:fig8.1.1a}
     \end{subfigure}
     \vskip -3pt
    \begin{subfigure}[b]{0.32\linewidth}
         \centering
         \includegraphics[width=0.99\linewidth]{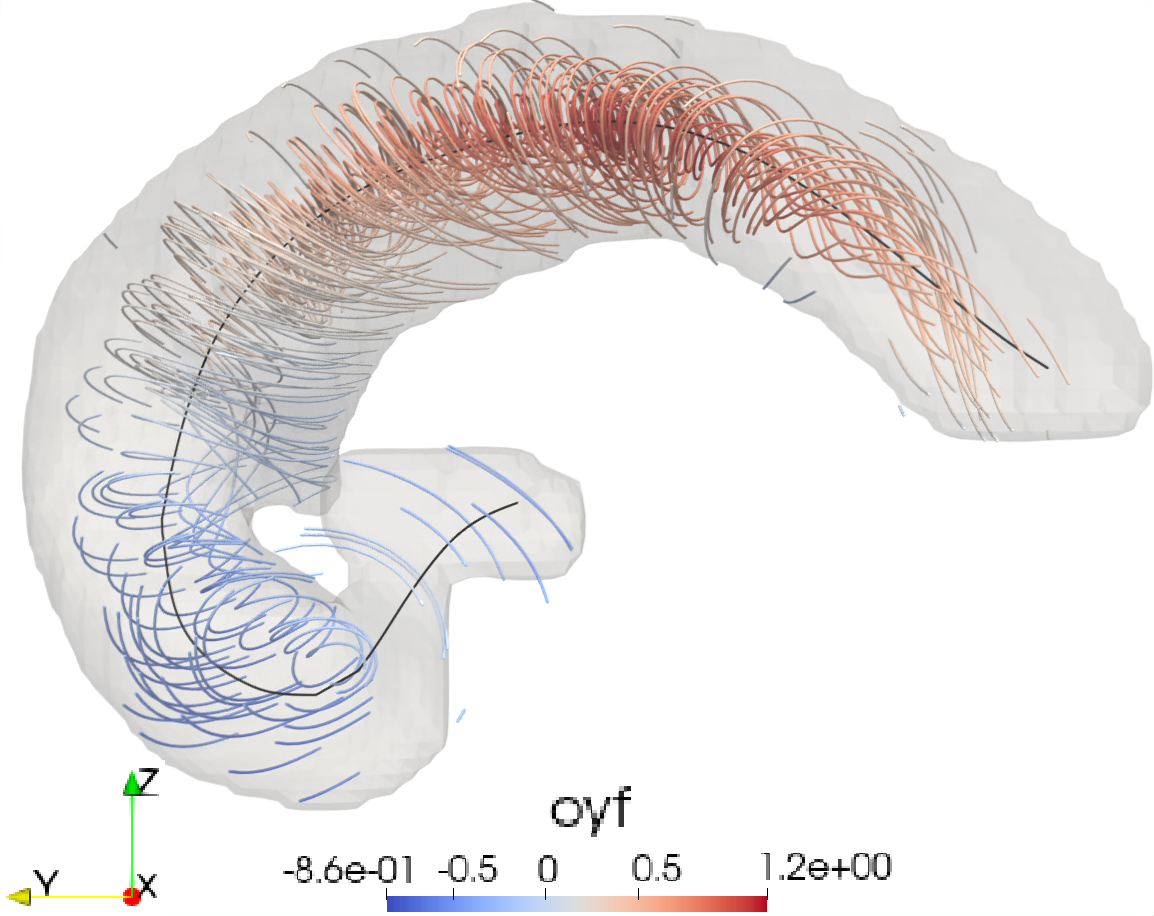}
         \vskip -2pt
         \caption{}
         \label{fig:fig8.1.1b}
     \end{subfigure}
     \begin{subfigure}[b]{0.32\linewidth}
         \centering
         \includegraphics[width=0.99\linewidth]{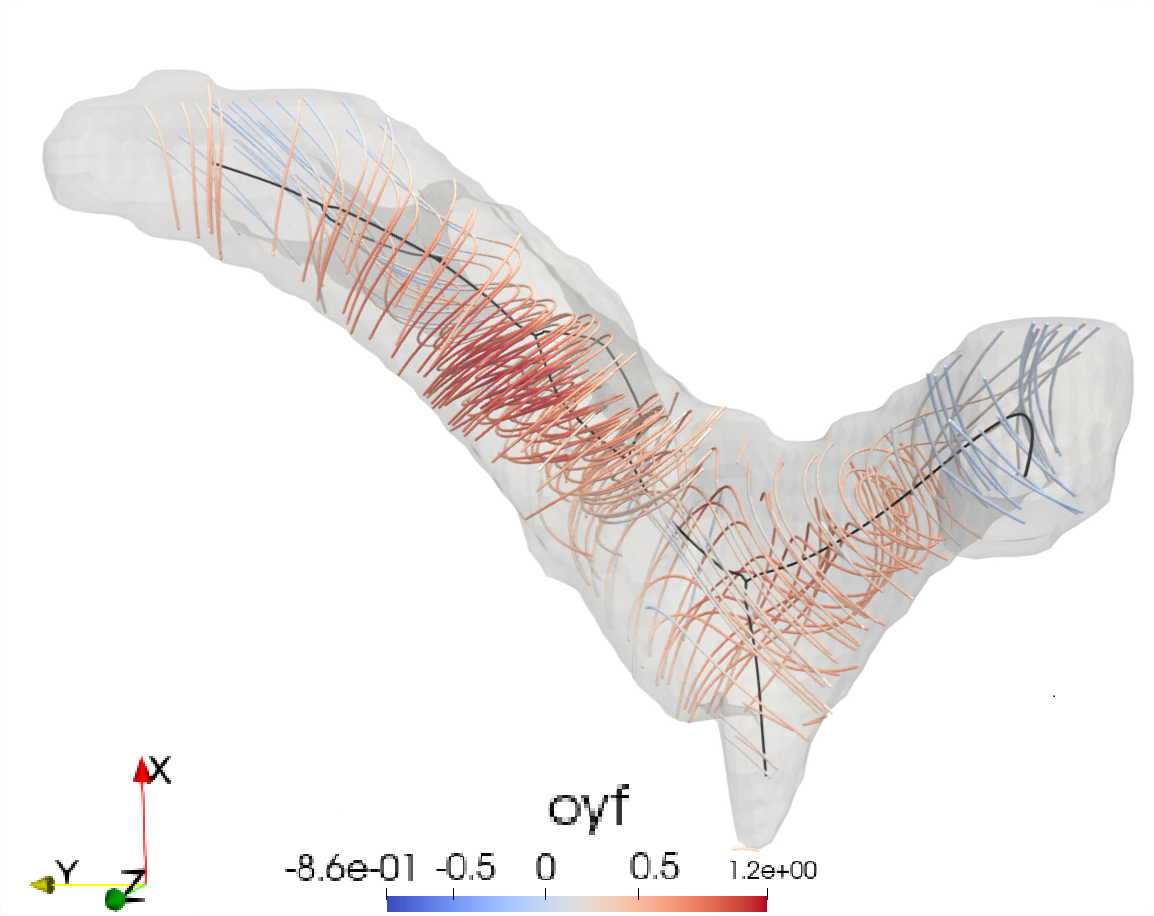}
         \vskip -2pt
         \caption{}
         \label{fig:fig8.1.1c}
     \end{subfigure}
    \begin{subfigure}[b]{0.32\linewidth}
         \centering
         \includegraphics[trim={0cm 0cm 0cm 0.5cm},clip,width=0.99\linewidth]{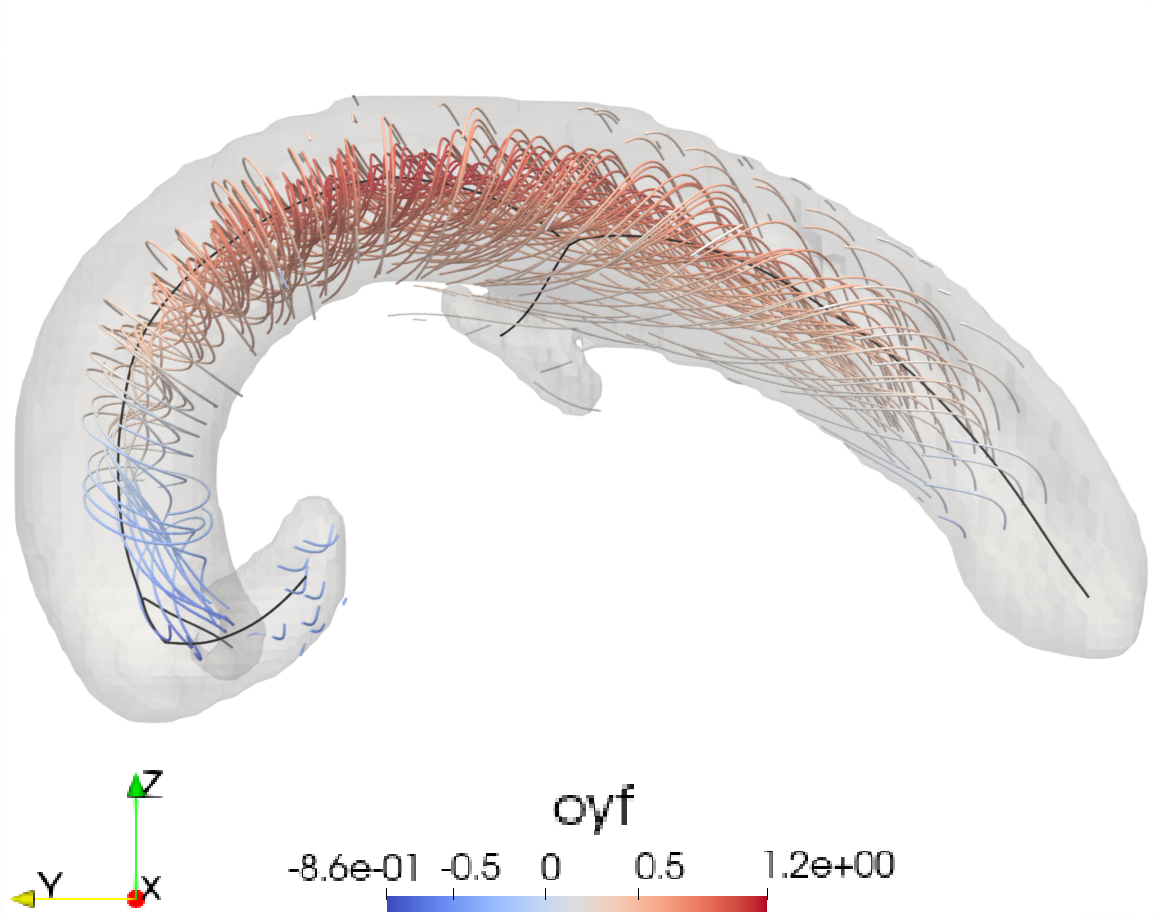}
         \vskip -2pt
         \caption{}
         \label{fig:fig8.1.1d}
    \end{subfigure}
    \vskip -3pt
    \begin{subfigure}[b]{0.99\linewidth}
         \centering
         \includegraphics[trim={0cm 0cm 0cm 0cm},clip,width=0.96\linewidth]{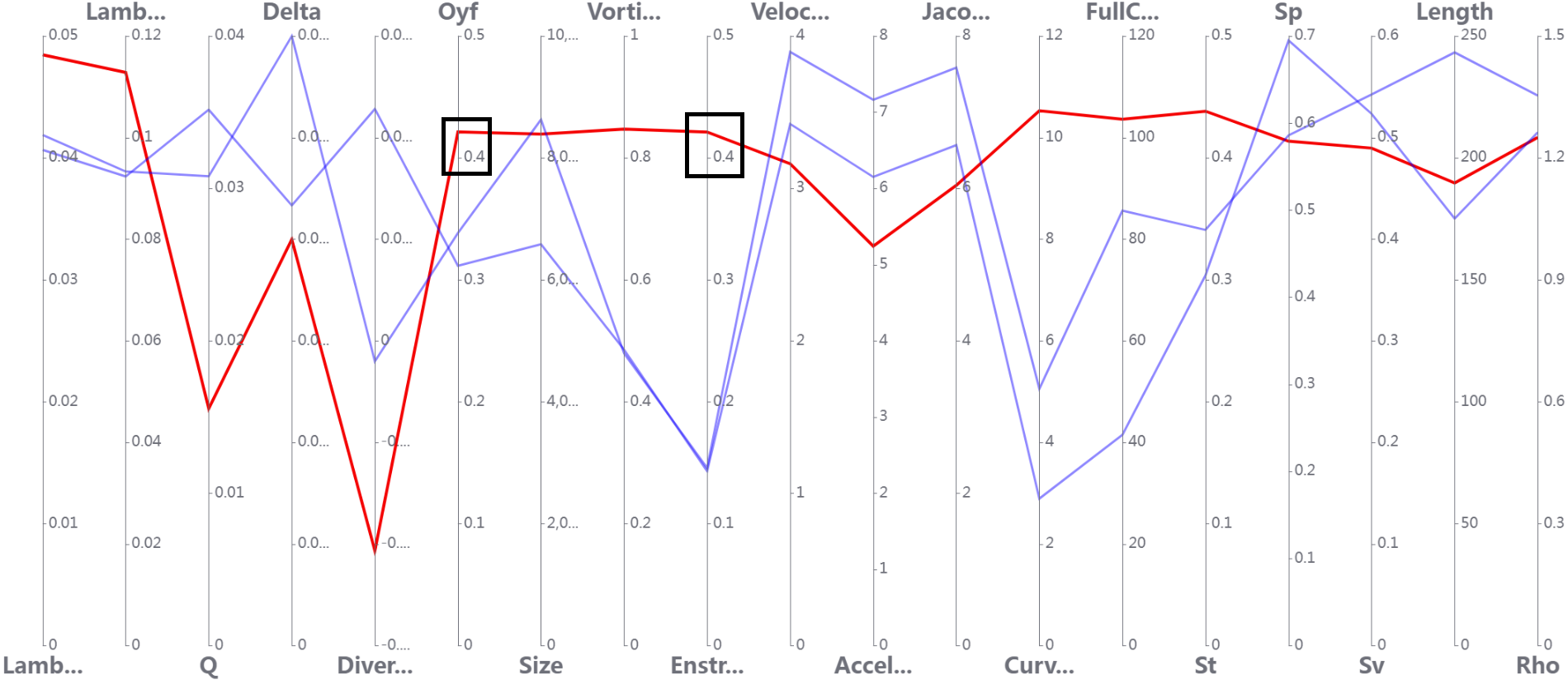}
         \vskip -2pt
         \caption{}
         \label{fig:fig8.1.1e}
     \end{subfigure}
     \vskip -2pt
 \subfigsCaption{
 Vortices of cluster 6 in \cref{fig:fig5.1a}. (a) highlights the vortices in the flow domain. Zoomed-in views of the three vortices from left to right are shown in (b), (c), and (d) along with their skeleton (black) and streamlines (colored by $\omega_y'$) (e) shows the PCP of the vortices.}
 \label{fig:fig8.1.1}
\end{figure}

\begin{figure}[!t]
 \centering 
      \begin{subfigure}[b]{0.99\linewidth}
         \centering
         \includegraphics[width=0.95\linewidth]{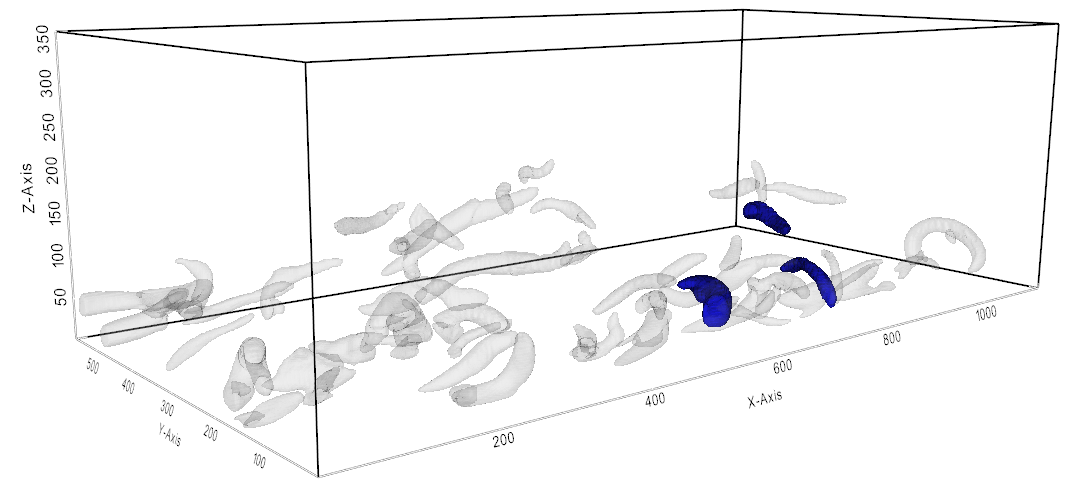}
         \vskip -5pt
         \caption{}
         \label{fig:fig8.1.2a}
     \end{subfigure}
     \vskip -2pt
    \begin{subfigure}[b]{0.32\linewidth}
         \centering
         \includegraphics[trim={0cm 0cm 0cm 8cm},clip,width=0.95\linewidth]{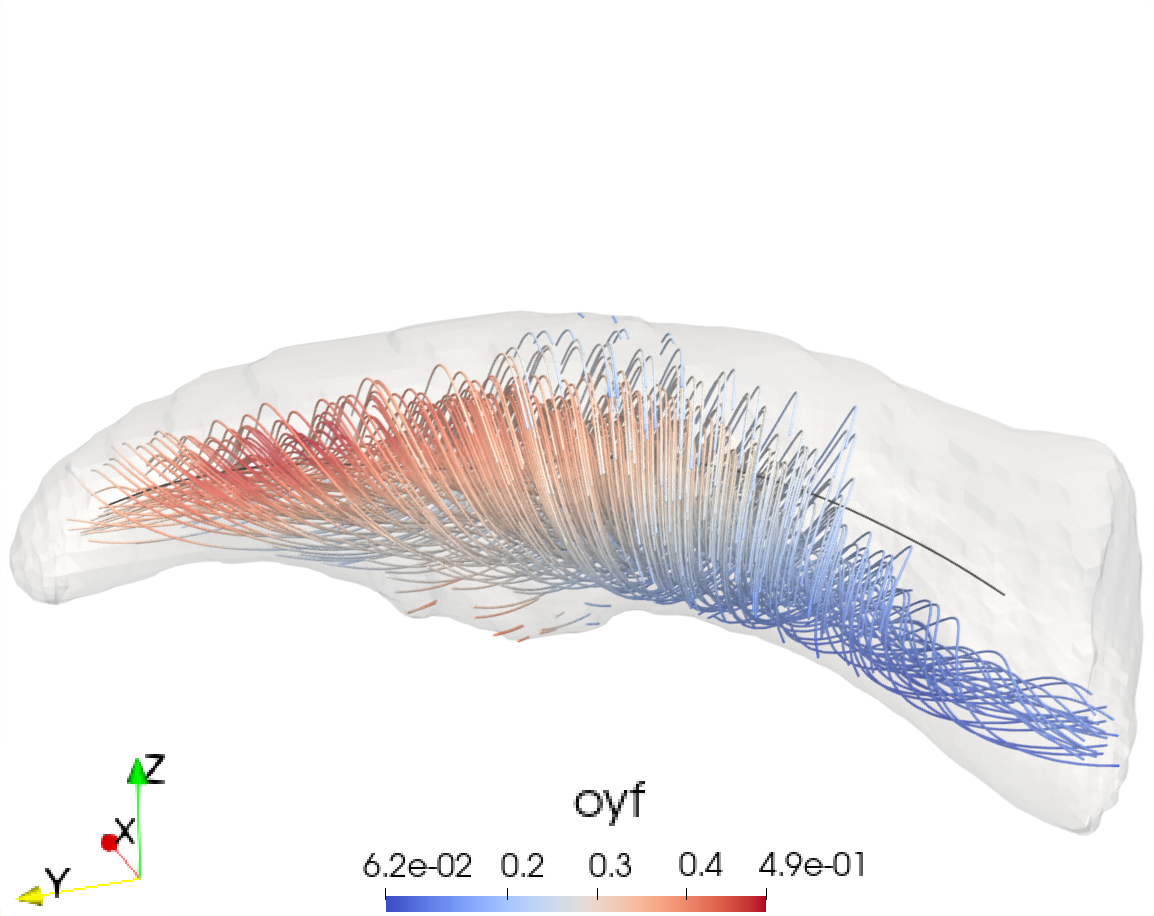}
         \vskip -2pt
         \caption{}
         \label{fig:fig8.1.2b}
     \end{subfigure}
     \begin{subfigure}[b]{0.32\linewidth}
         \centering
         \includegraphics[trim={0cm 0cm 0cm 10cm},clip,width=0.95\linewidth]{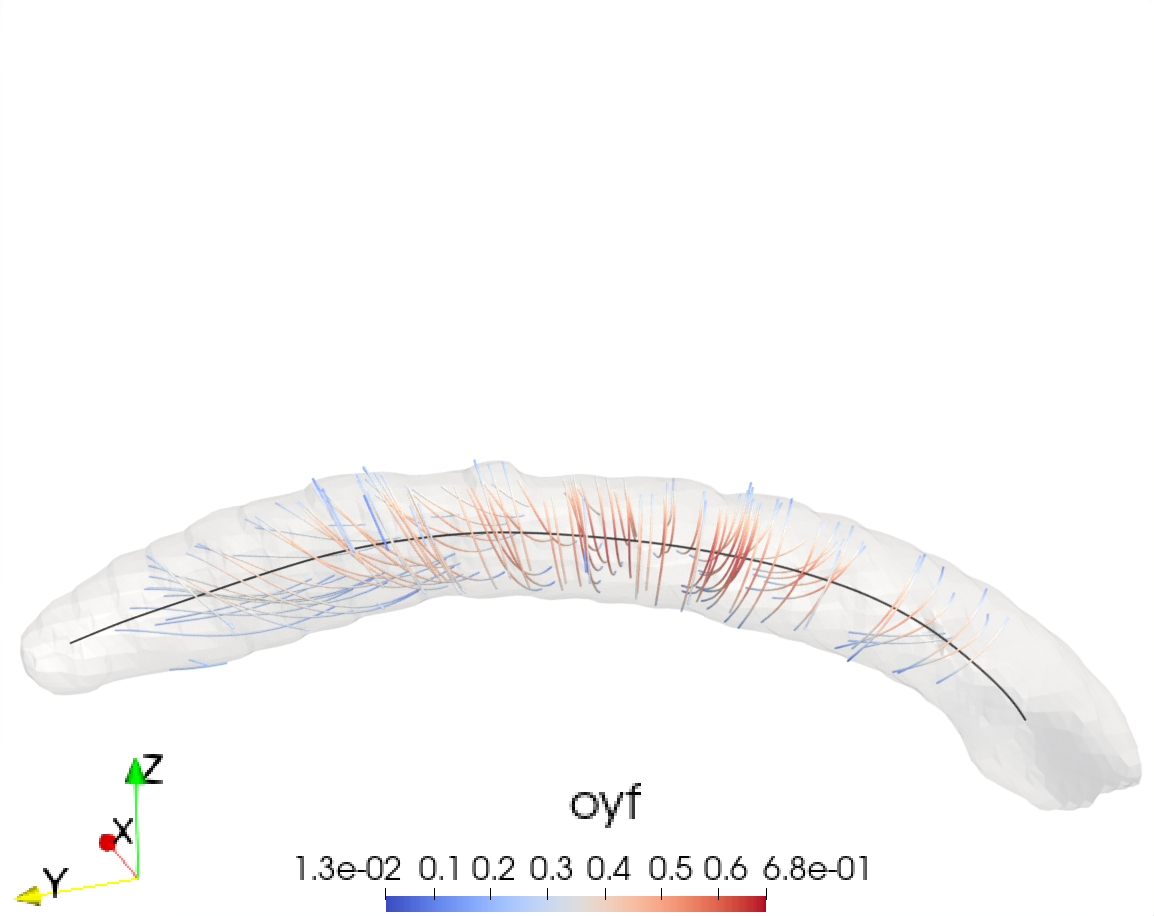}
         \vskip -2pt
         \caption{}
         \label{fig:fig8.1.2c}
     \end{subfigure}
    \begin{subfigure}[b]{0.32\linewidth}
         \centering
         \includegraphics[trim={0cm 0cm 0cm 10cm},clip,width=0.95\linewidth]{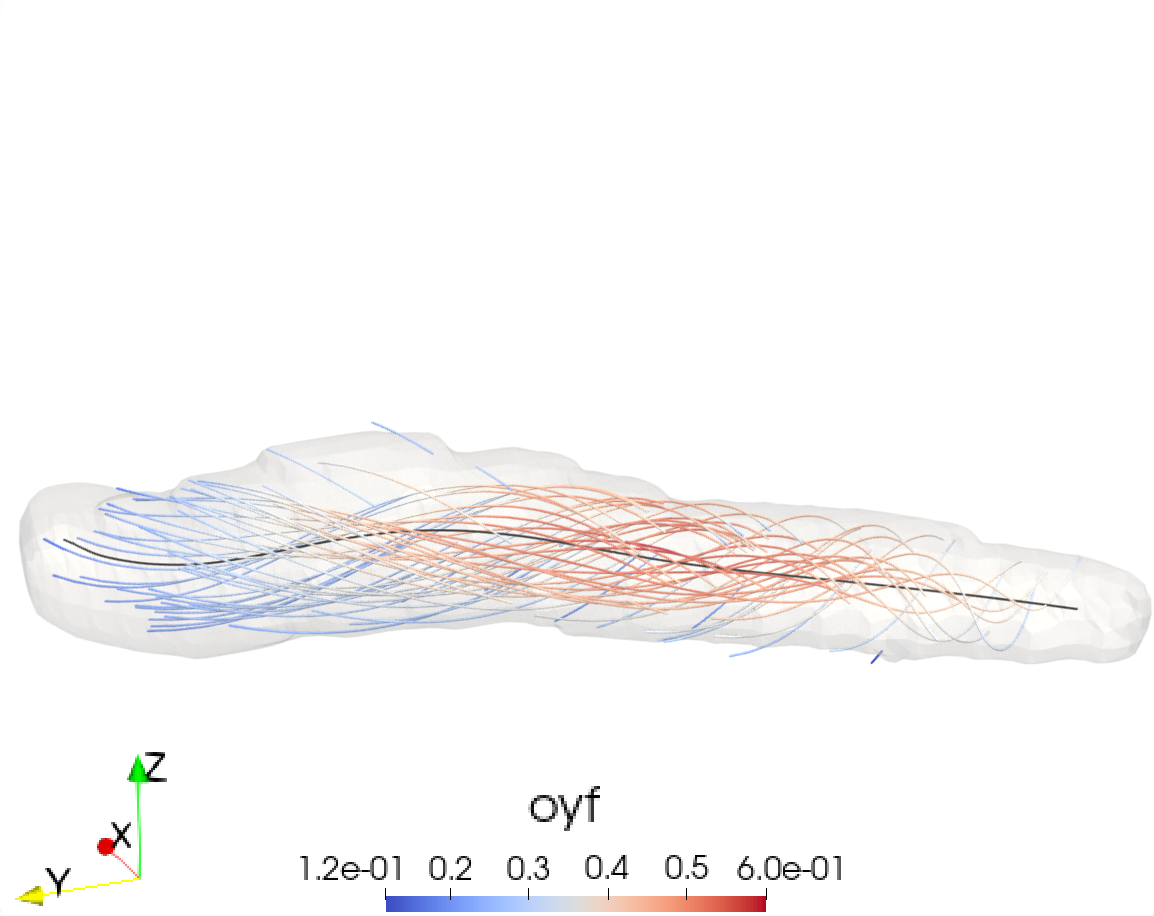}
         \vskip -2pt
         \caption{}
         \label{fig:fig8.1.2d}
    \end{subfigure}
    \vskip -2pt
    \begin{subfigure}[b]{0.99\linewidth}
         \centering
         \includegraphics[width=0.96\linewidth]{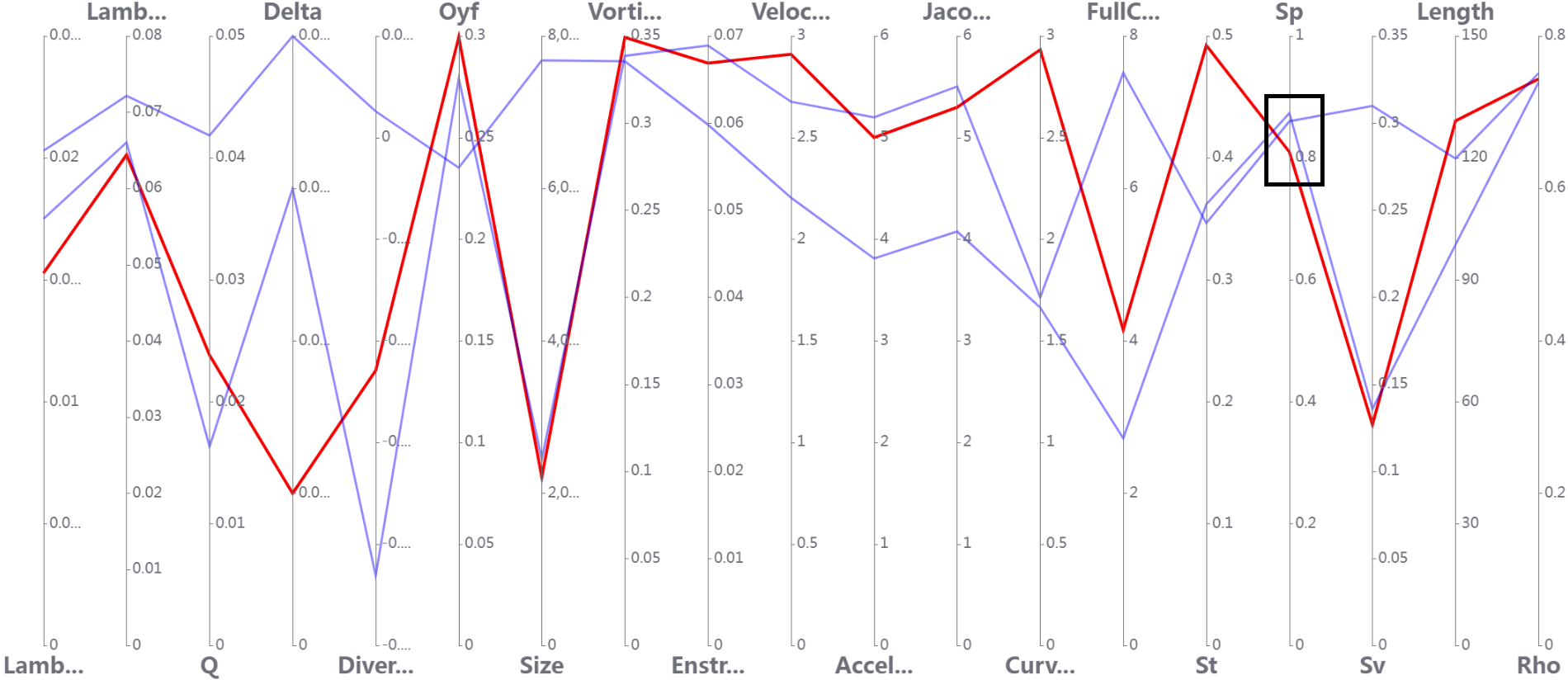}
         \vskip -2pt
         \caption{}
         \label{fig:fig8.1.2e}
     \end{subfigure}
     \vskip -2pt
 \subfigsCaption{This figure shows vortices belonging to the cluster 5 in \cref{fig:fig5.1a}. 3D View (a) highlights the vortices in the flow domain. Zoomed-in versions of the bottom left, bottom right, and top vortex are shown in (b), (c), and (d) respectively, along with their skeleton (black) and streamlines (colored by $\omega_y'$) (e) shows the PCP of the vortices.}
 \label{fig:fig8.1.2}
\end{figure}

\begin{figure}[!t]
 \centering 
      \begin{subfigure}[b]{0.99\linewidth}
         \centering
         \includegraphics[width=0.97\linewidth]{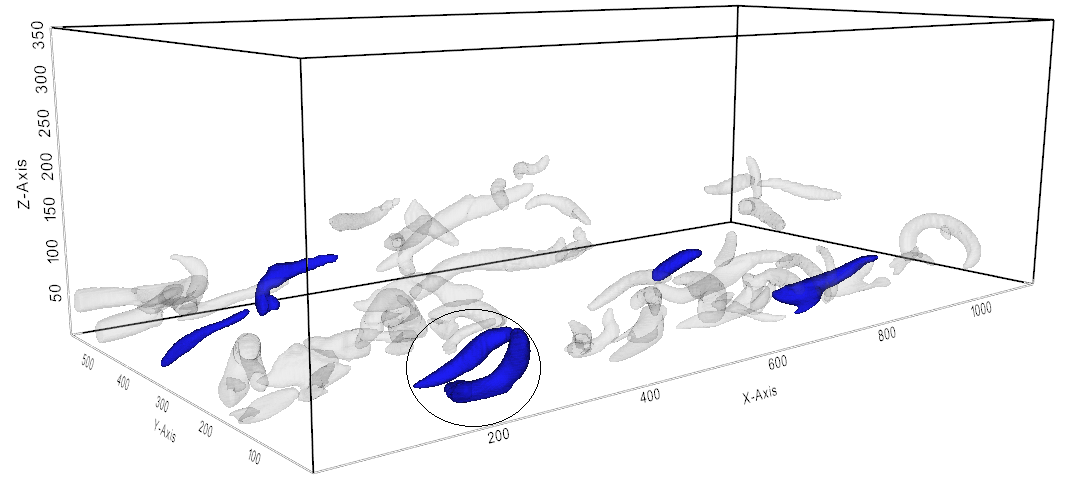}
         \vskip -5pt
         \caption{}
     \end{subfigure}
     \vskip -5pt
    \begin{subfigure}[b]{0.99\linewidth}
         \centering
         \includegraphics[width=0.99\linewidth]{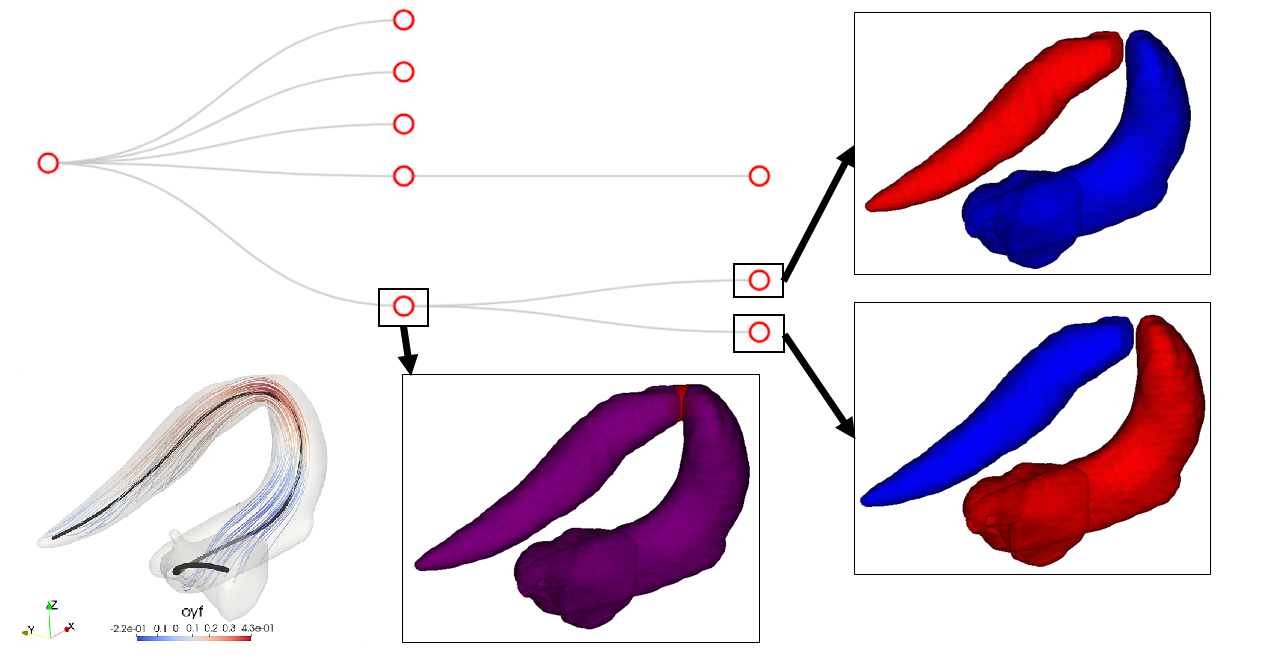}
         \vskip -5pt
         \caption{}
         \label{fig:fig8.1.3b}
     \end{subfigure}
    \vskip -5pt
    \begin{subfigure}[b]{0.99\linewidth}
         \centering
         \includegraphics[width=0.99\linewidth]{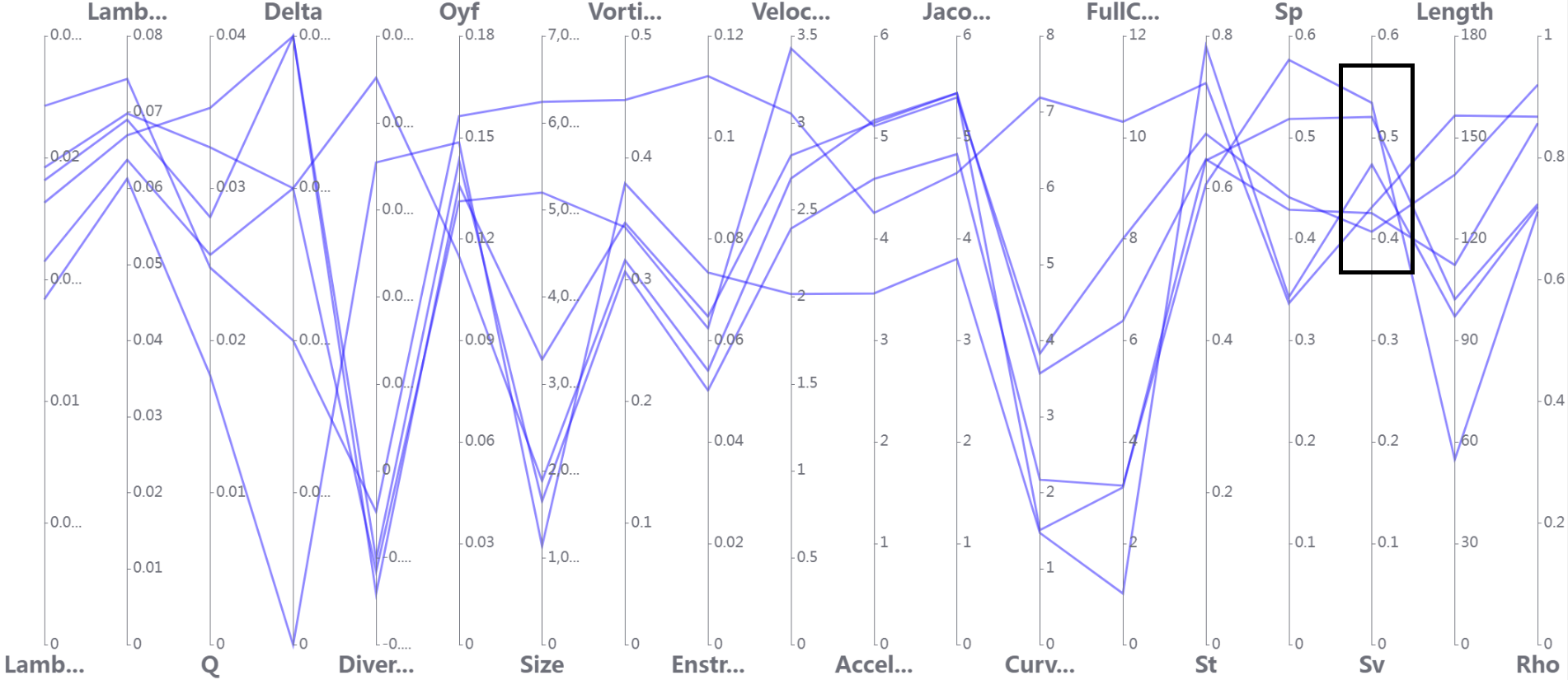}
         \caption{}
     \vskip -2pt
     \end{subfigure}
     \vskip -2pt
 \subfigsCaption{
 Vortices of cluster 3 in \cref{fig:fig5.1a}. (a) highlights the vortices in the flow domain. (b) shows the tree view and hierarchy of two vortices highlighted in a circle in (a). The vorticity lines in the leftmost vortex in (b) are color-coded with $\omega_y'$. (e) shows the PCP of the vortices.
 }
 \label{fig:fig8.1.3}
\end{figure}

\begin{figure}[t]
 \centering 
      \begin{subfigure}[b]{0.99\linewidth}
         \centering
         \includegraphics[width=0.97\linewidth]{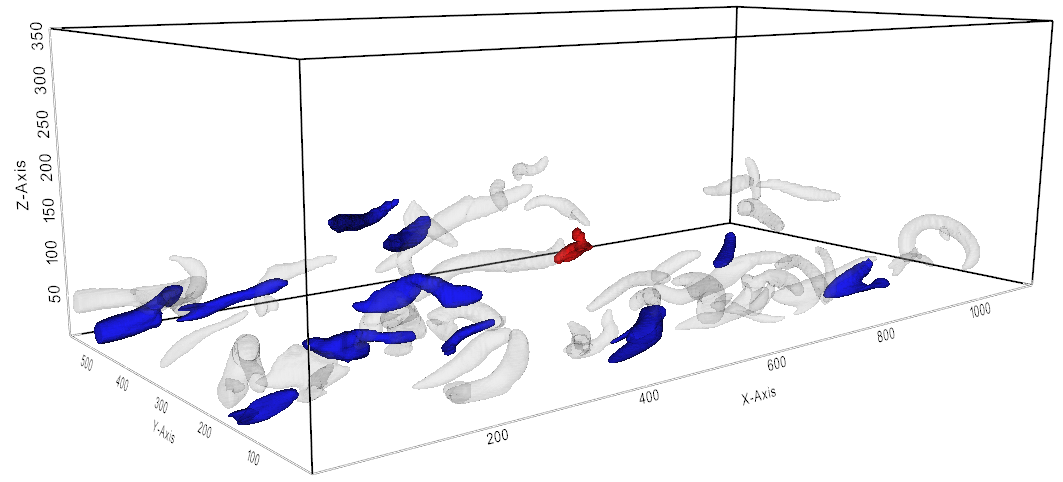}
         \vskip -10pt
         \caption{}
         \label{fig:fig8.1.4a}
     \end{subfigure}
     \vskip -10pt
    \begin{subfigure}[b]{0.32\linewidth}
         \centering
         \includegraphics[trim={0cm 0cm 0cm 4cm},clip,width=0.99\linewidth]{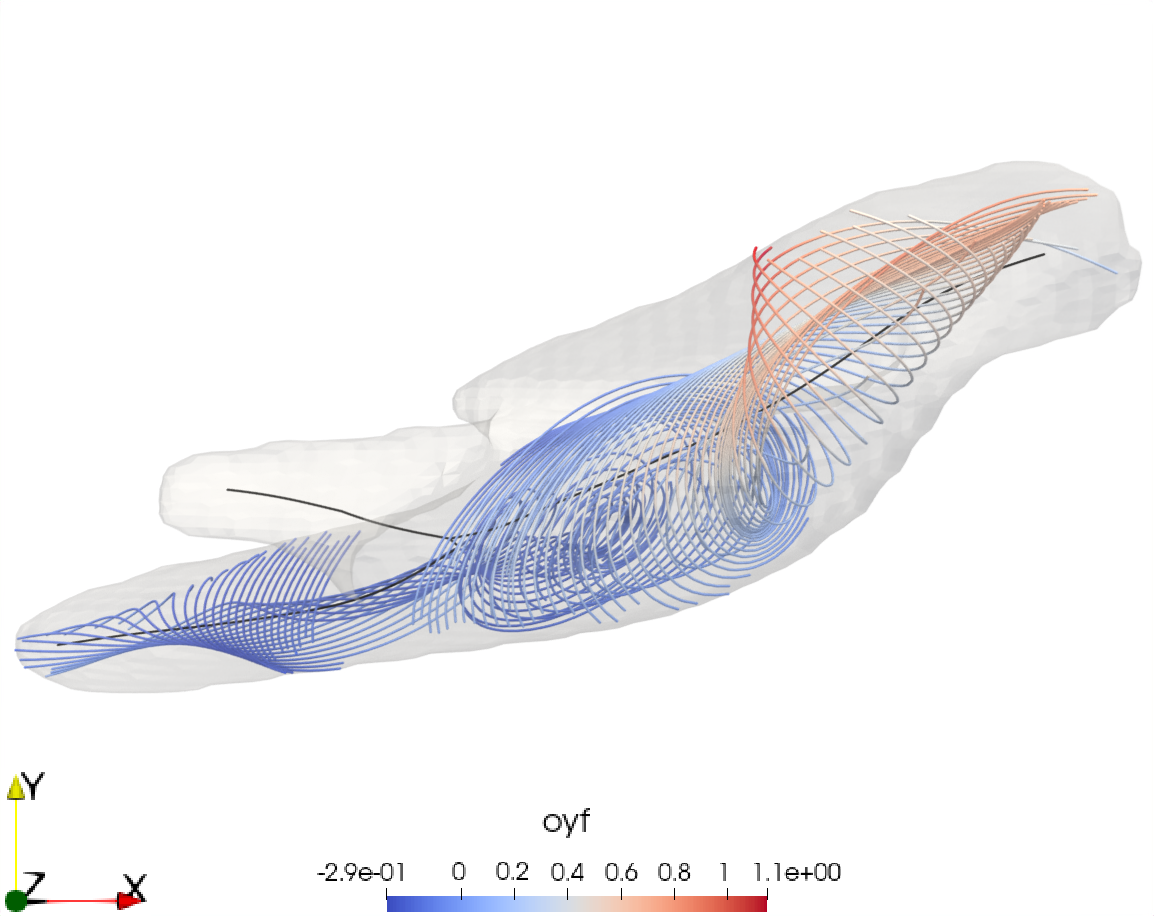}
         \vskip -2pt
         \caption{}
         \label{fig:fig8.1.4b}
     \end{subfigure}
     \begin{subfigure}[b]{0.32\linewidth}
         \centering
         \includegraphics[trim={0cm 0cm 0cm 4cm},clip,width=0.99\linewidth]{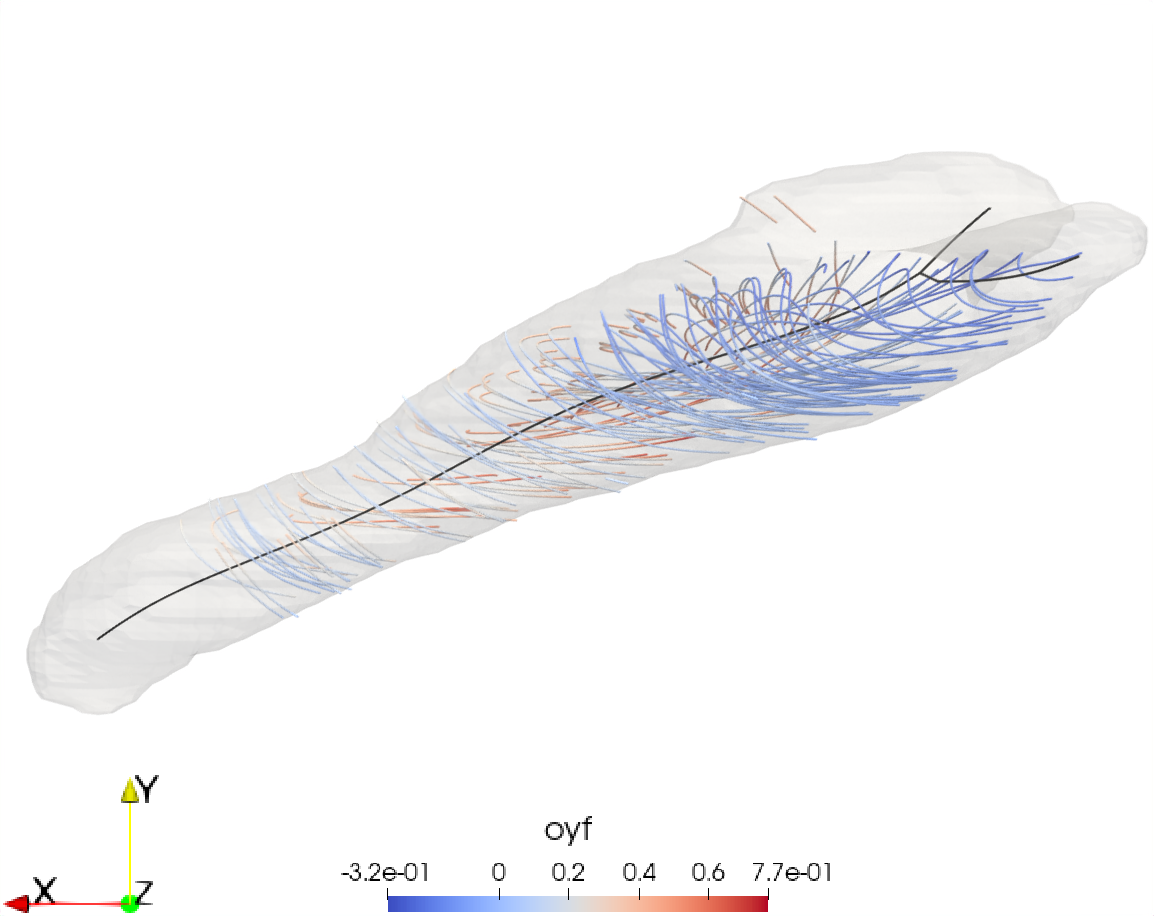}
         \vskip -2pt
         \caption{}
         \label{fig:fig8.1.4c}
     \end{subfigure}
    \begin{subfigure}[b]{0.32\linewidth}
         \centering
         \includegraphics[trim={0cm 0cm 0cm 0cm},clip,width=0.99\linewidth]{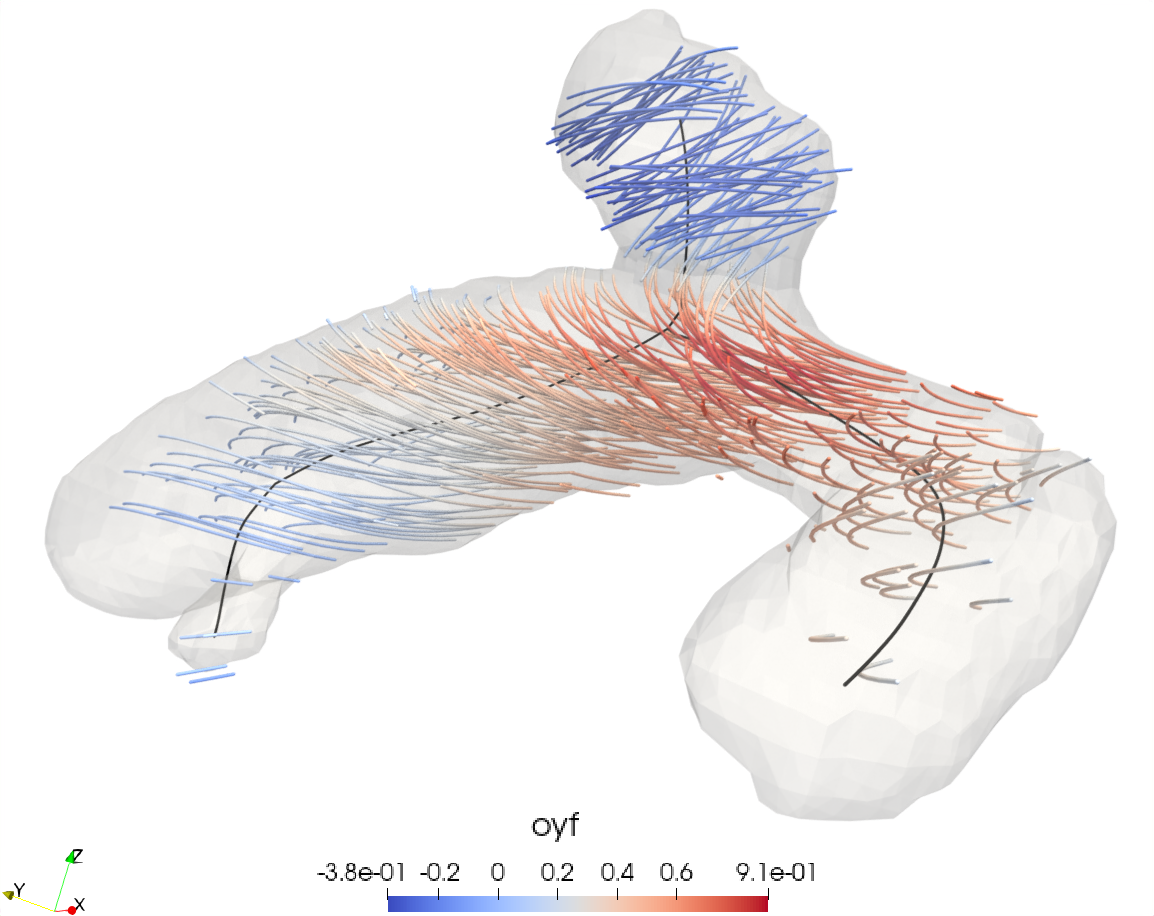}
         \vskip -2pt
         \caption{}
         \label{fig:fig8.1.4d}
    \end{subfigure}
    \vskip -5pt
    \begin{subfigure}[b]{0.99\linewidth}
         \centering
         \includegraphics[width=0.99\linewidth]{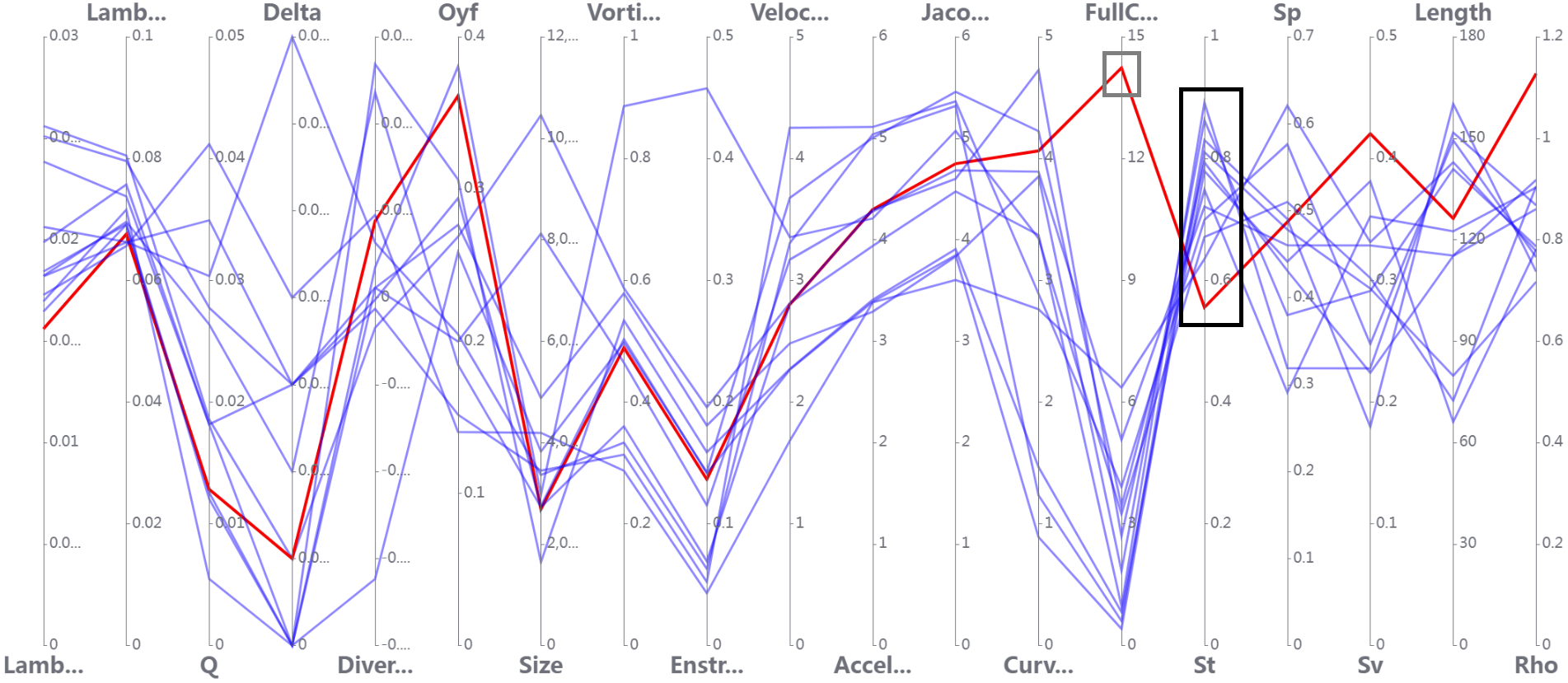}
         \caption{}
         \label{fig:fig8.1.4e}
     \end{subfigure}
     \vskip -2pt
 \subfigsCaption{
 Vortices of cluster 0 in \cref{fig:fig5.1a}. 3D View (a) highlights the vortices in the flow domain. Zoomed-in versions of the top three nodes of the tree view are shown in (b), (c), and (d), respectively, along with their skeleton (black) and streamlines (colored by $\omega_y'$). (e) shows the PCP of the vortices. (b) and (c) show the top-down view and (d) shows the side angle view.
 }
 \label{fig:fig8.1.4}
\end{figure}

\Cref{fig:fig8.1.1} shows vortices belonging to cluster 6 from the clustering results of \cref{fig:fig5.1a}. This cluster reveals two ``Omega'' shaped structures that best depict the geometry of a hairpin. These ``Omega'' shaped hairpins usually are in their later stage (i.e., disconnected from their legs) and will be lifted to the middle part of the flow. The streamlines show strong vorticity and the presence of strong positive $\omega_y'$ in the curved sections of the vortices resembling the physical and geometric characteristics of the head of a hairpin vortex as discussed in \cref{sec:hairpin}. The vortex shown in \cref{fig:fig8.1.1c} doesn't have a perfect "omega" like shape but does have a strong positive $\omega_y'$ and enstrophy as shown in the highlighted line (red) in the PCP. It could consist of a part of a hairpin (left branch) and another non-hairpin vortex (right branch).

\Cref{fig:fig8.1.2} shows vortices belonging to cluster 5. This cluster reveals “arches” with strong positive $\omega_y'$ in the spanwise direction depicted by the high $S_p$ component highlighted (black box) in the PCP view. This "arch" like shape in the spanwise direction is more aligned with stage (a) of the life span. The value of $\omega_y'$ in the vortex in \cref{fig:fig8.1.2b} shows a decreasing trend from left to right which is visible by the color of streamlines. This pattern of $\omega_y'$ is similar to the pattern of one-legged hairpin as discussed in \cref{sec:hairpin}.

\Cref{fig:fig8.1.3} shows vortices belonging to cluster 3. This cluster reveals vortices having a high roll-up/elevation which is depicted by the grouping of PCP lines at the $S_v$ component of the PCP (highlighted in black). However, most of them are not hairpin vortices at their current stage, except for the ones highlighted. The close-up view (\cref{fig:fig8.1.3b}) of these two vortices show they assemble the two legs of a hairpin vortex with a high elevation toward the streamwise direction. 
\Cref{fig:fig8.1.3b} also shows an example of the usability of the tree view. The leaf nodes highlighted in black correspond to two vortices highlighted in red in their respective figures. Upon selection of the parent node of these vortices, it is revealed that the two vortices actually belong to the same vortex which has a shape like a hairpin and has a strong positive value of $\omega_y'$ close to the head. The contiguous skeleton (black) and the continuous vorticity lines verify the fact that it is the same vortex that suffered from a degenerate split.

\Cref{fig:fig8.1.4} shows the vortices belonging to cluster 0. The vortices in this cluster are mostly in the streamwise direction as depicted by the grouping of lines at $S_t$ axis (highlighted in the black box) in the PCP view (\cref{fig:fig8.1.4e}) and the direction of the vortices in \cref{fig:fig8.1.4b} and \cref{fig:fig8.1.4c}. Most of them are not hairpin vortices. However, the expert points out that the vortex in \cref{fig:fig8.1.4b} could be an incomplete hairpin vortex due to its uplifting orientation and strong spanwise vorticity at the right end, which is also depicted by the well-organized swirling pattern of streamlines. The vortex in \cref{fig:fig8.1.4d}, has a different symmetry than other vortices in the cluster which is depicted by the curvature axis (highlighted in the grey box) of the highlighted (red) line of the PCP plot. It has a much high curvature but low $S_t$, compared to the rest in cluster 0. It could be a part of a hairpin, but probably less organized based on the pattern shown by the streamlines.



\noindent \textbf{Expert Evaluation.} The system and results have been evaluated by the expert. The expert made a general comment, saying that the proposed method allows them ``\emph{to observe a significant variation of the detailed characteristics (e.g., geometry, size, elevation, orientation, etc.) of each identified structure, including hairpin vortices.  Based on these individual samples, we can obtain the true statistics of a certain type of structure.}'' Among the identified candidate hairpin vortices, ``\emph{clearly, some of them are hairpin vortices, but some of them are not.  This is actually great.''. Some of these vortices, while do not look like a hairpin, ``may still be related to hairpin vortices}''. This is because ``...\emph{all the flow structures are evolving continuously in a turbulent flow field due to disturbances from other neighboring flow structures.  A hairpin vortex may be generated by some instability mechanism or by flow induction of an existing vortex structure.  It may start with the head first or the two legs first; evolve into a full hairpin; with one leg destroyed by other neighboring flow structures; eventually evolve back to a one-legged quasi-streamwise/spanwise vortex or a pure spanwise vortex with the head only}''.  The expert also pointed out that to further investigate whether some possible candidates (e.g., the vortex shown in \cref{fig:fig8.1.4b}) are hairpins or not, ''\emph{we will need to look at a series of instantaneous snapshots in a short time period to identify and track individual vortex structures}'' in the future. Nonetheless, the expert believes the developed technique and system provide a valuable and unified tool for the study of hairpins and other vortices in turbulent flows.

\subsection{Application to \revise{Other Flows}}
\label{section8.1}
We next apply our method to other flows, including the B\'enard flow, the flow behind a square cylinder, Plume, and Crayfish. Due to limited space, we only present the results for B\'enard and Cylinder flows.


\begin{figure}[!t]
 \centering 
      \begin{subfigure}[b]{0.49\linewidth}
         \centering
         \includegraphics[width=0.96\linewidth]{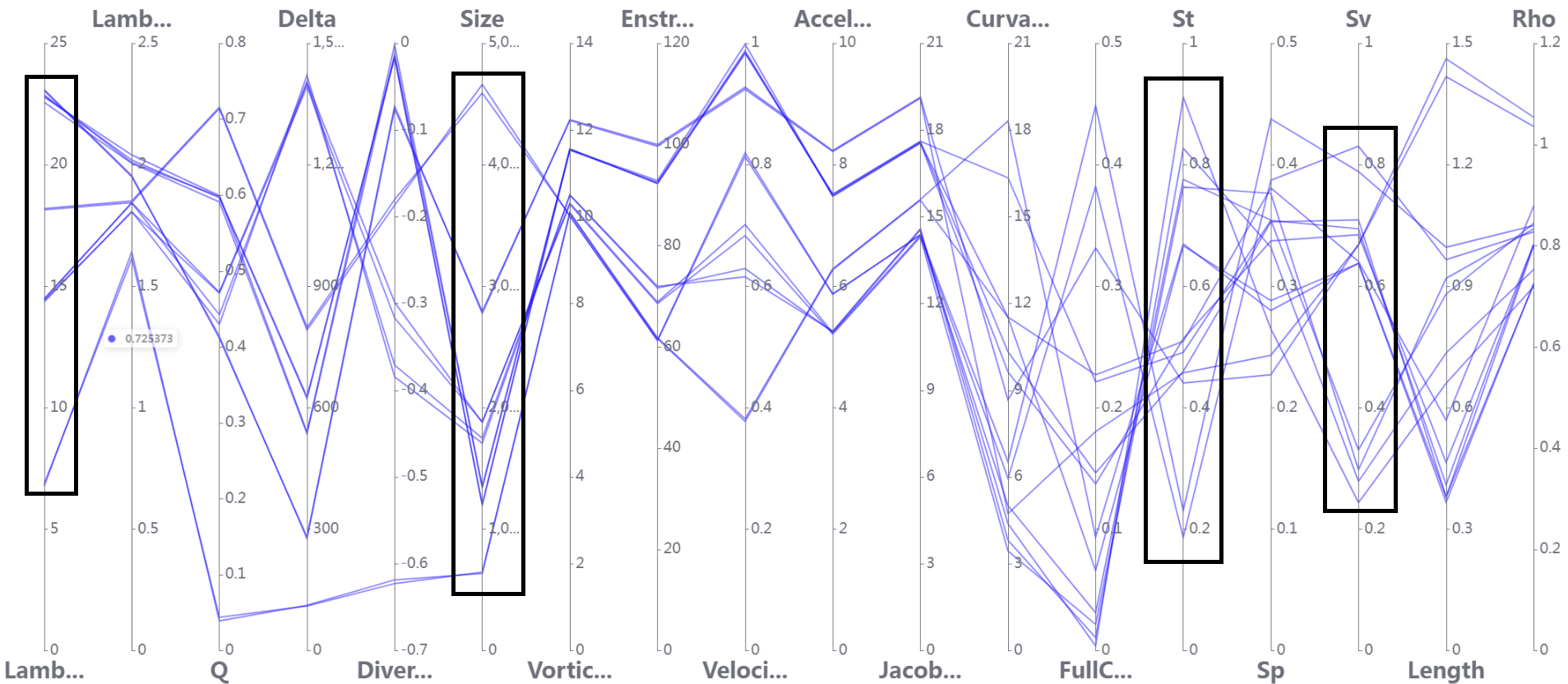}
         \caption{}
         \label{fig:fig8.2.1a}
     \end{subfigure}
           \begin{subfigure}[b]{0.49\linewidth}
         \centering
         \includegraphics[width=0.96\linewidth]{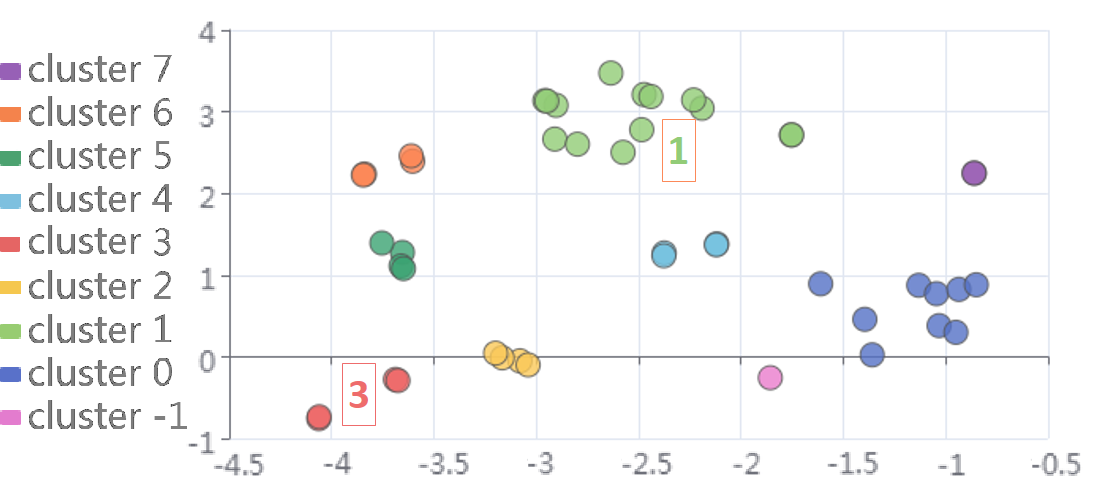}
         \caption{}
         \label{fig:fig8.2.1b}
     \end{subfigure}
     \vskip -3pt
     \begin{subfigure}[b]{0.49\linewidth}
         \centering
         \includegraphics[width=0.96\linewidth]{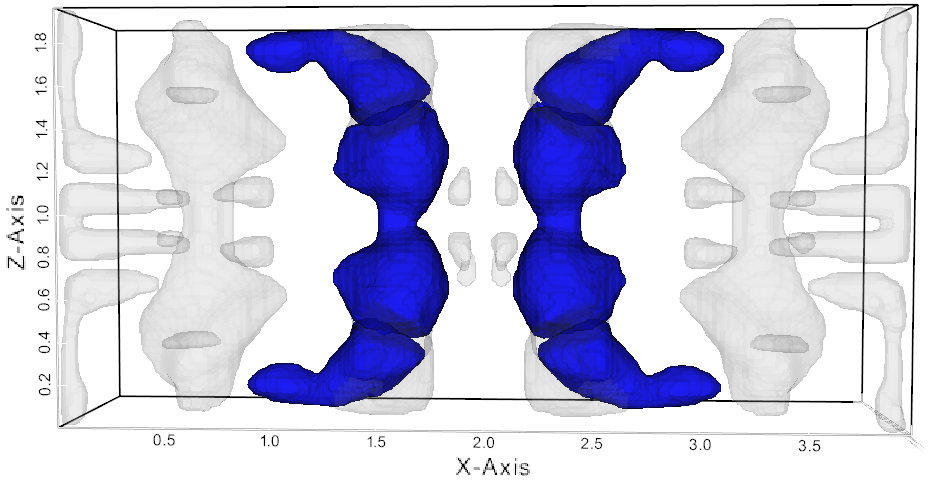}
         \caption{}
         \label{fig:fig8.2.1c}
     \end{subfigure}
          \begin{subfigure}[b]{0.49\linewidth}
         \centering
         \includegraphics[width=0.96\linewidth]{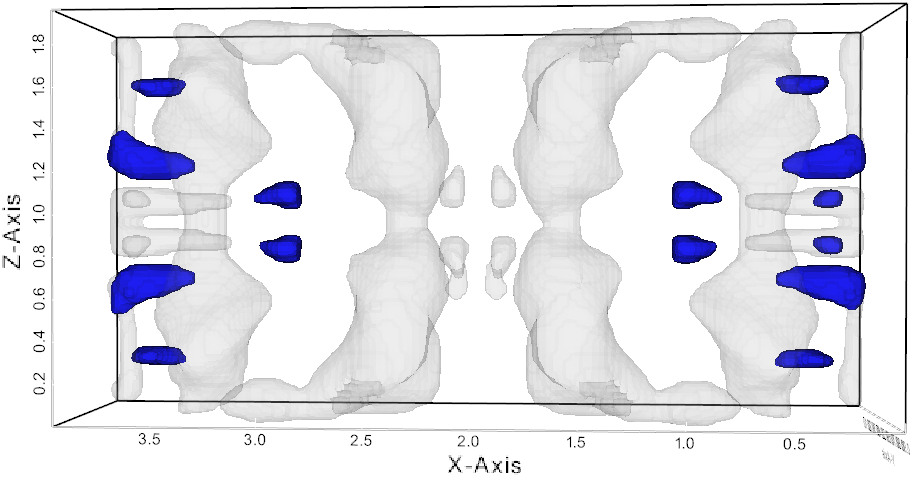}
         \caption{}
         \label{fig:fig8.2.1d}
     \end{subfigure}
     \vskip -4pt
 \subfigsCaption{Clustering results for the B\'enard flow. (a) shows the PCP view and (b) shows the scatter plot of the clustering results. (c) and (d) show the vortices corresponding to clusters 3 and 1, respectively.
 }
 \label{fig:fig8.2.1}
\end{figure}

\Cref{fig:fig8.2.1} shows the clustering results for the B\'enard flow. First, we analyze the PCP (\cref{fig:fig8.2.1a}) for all vortices to shortlist the features/attributes to use for clustering. The features belonging to the axis of the PCP having a high variance/spread are better suited to use for clustering (as highlighted in black). Depending on the analysis of the PCP view, we use $\lambda_2$, $size$, $S_t$, and $S_v$ for clustering. Results reveal the vortices of different scales and orientations in \cref{fig:fig8.2.1c} and \cref{fig:fig8.2.1d}. Especially, the vortices in \cref{fig:fig8.2.1c} correspond to two primary convection vortices, while the smaller vortices in \cref{fig:fig8.2.1d} are the secondary vortices with orientation perpendicular to the primary vortices. The clustering captures the symmetry properties of these vortices and separates them based on their scales and orientations.

\begin{figure}[t]
 \centering 
      \begin{subfigure}[b]{0.99\linewidth}
         \centering
         \includegraphics[width=0.95\linewidth]{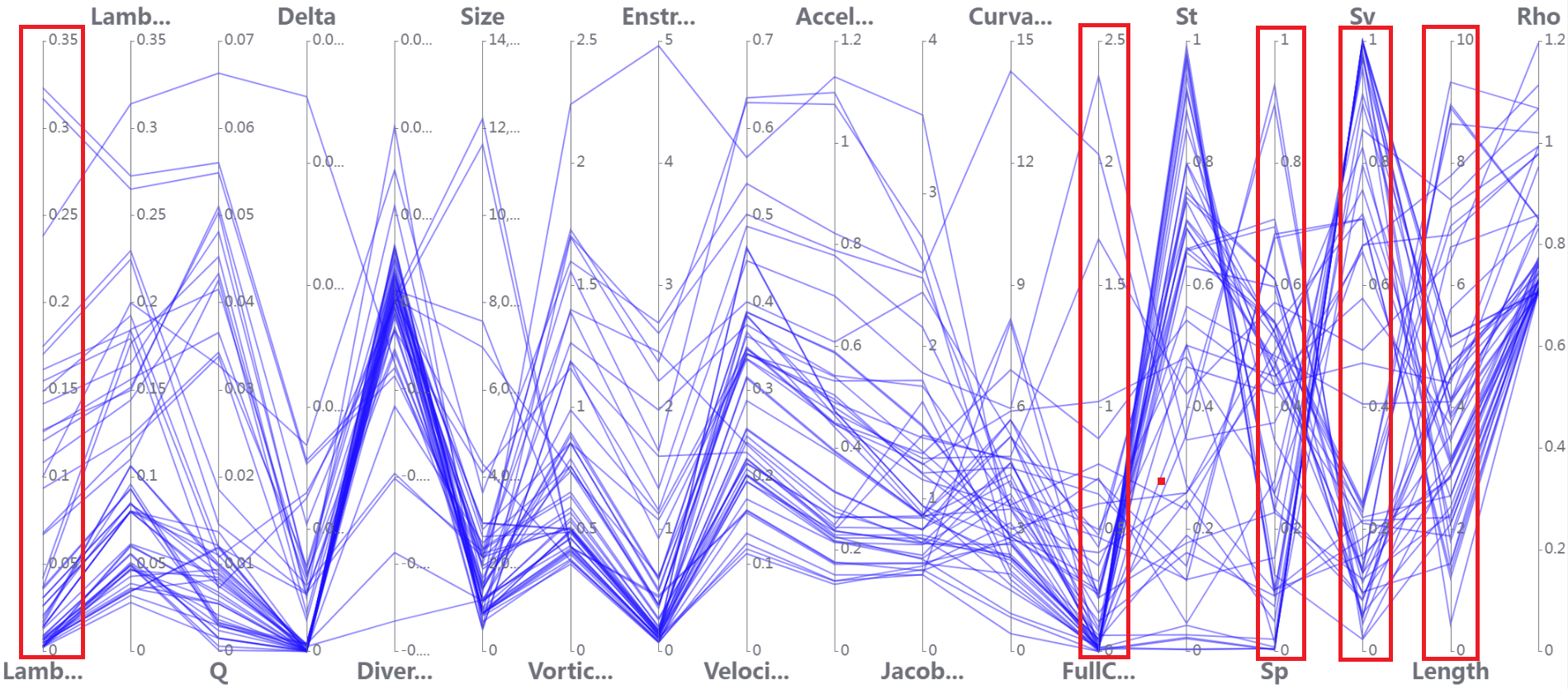}
         \caption{}
         \label{fig:fig8.2.2a}
     \end{subfigure}
     \vskip -3pt
    \begin{subfigure}[b]{0.99\linewidth}
         \centering
         \includegraphics[width=0.95\linewidth]{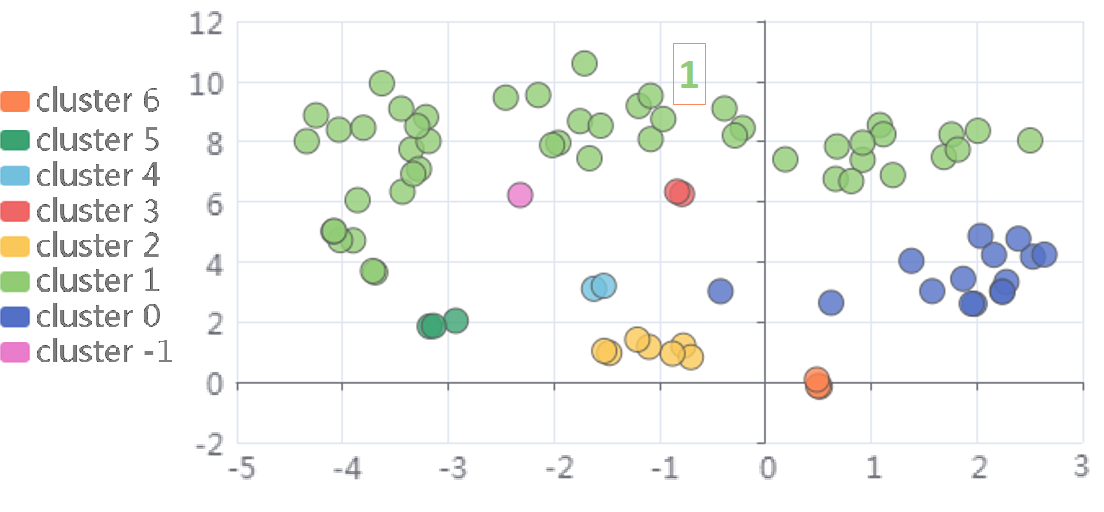}
         \caption{}
         \label{fig:fig8.2.2b}
     \end{subfigure}
     \vskip -3pt
     \begin{subfigure}[b]{0.99\linewidth}
         \centering
         \includegraphics[width=0.95\linewidth]{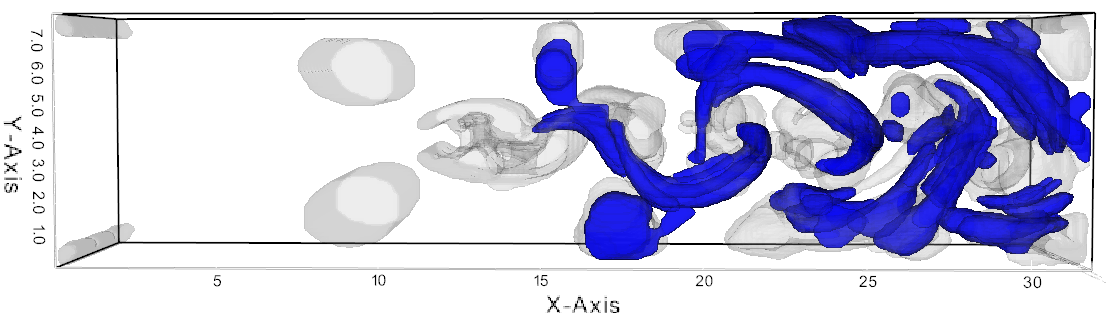}
         \caption{}
         \label{fig:fig8.2.2c}
     \end{subfigure}
     \vskip -2pt
 \subfigsCaption{Clustering results for the Cylinder flow. (a) shows the PCP view and (b) shows the scatter plot of the clustering results. (c) shows the vortices corresponding to cluster 1 in (b).
 }
 \label{fig:fig8.2.2}
\end{figure}

We use the same approach to shortlist features to cluster the vortices in the Cylinder Flow. The results are shown in \cref{fig:fig8.2.2}. We use the PCP (\cref{fig:fig8.2.2a}) to select the features with a large spread (highlighted in red) to use for clustering. All the vortices in the von K\'arm\'an vortex street belong to cluster 1 in \cref{fig:fig8.2.2b} which are highlighted in \cref{fig:fig8.2.2c}. This shows a powerful demonstration of clustering the vortices based on the combination of physical and geometric attributes.



\section{Summary and Future Work}
\label{sec:conclusion}

In this work, we introduced a framework for hairpin vortex extraction and characterization. 
Our framework extracts the individual vortices using a region-growing and region-splitting process. A profile is then built for each vortex using its relevant physical and geometric attributes. Next, a set of criteria are defined and applied to select candidate hairpin vortices, from which the true hairpin vortices are separated using clustering. We develop a visualization system to support the user exploration and analysis of the extracted vortices using different linked-views. We demonstrate the effectiveness of our method and system by extracting hairpin vortices from the stress-driven turbulent Couette flow and by applying it to the exploration of vortices in other flows. 



\textbf{Limitations:} Degenerate vortex splitting is the major limitation in the current vortex separation approach. It limits the ability to perform geometric analysis on the separated vortices. In addition, the current filtering criteria for identifying hairpin vortices are still not sufficiently refined, leading to candidates with some false positive hairpins. This is similar to the identification of other types of vortices due to the lack of knowledge of those vortices. 
\revise{Finally, our method doesn't generalize to turbulence without predominant streamwise direction (e.g. homogeneous isotropic turbulence)}.

\textbf{Future Work:} We plan to address the limitations mentioned above. \revise{In particular, we will look into \cite{schneider2008interactive}, they tried to solve the vortex splitting problem using the largest contour tree segmentation of the contour tree}. In addition, we will extend our framework to handle the time-dependent turbulent flows for vortex tracking. We will also apply it to the slip boundary (i.e., the top boundary of the Couette flow) where the knowledge about hairpin vortices there is little.


\section*{Supplemental Materials}
\label{sec:supplemental_materials}
The supplemental materials include (1) a document containing details of the physical attributes mentioned in \cref{Tab:table1}, the datasets used in \cref{tab:Table2} and a figure explaining stages of the hairpin vortices, (2) a 
video explaining the usage of the interactive visualization system (\cref{sec:visualizationsystem}) and (3) the \href{https://github.com/adeelz92/hairpin-vortices-extraction.git}{CODE} for vortex extraction and separation (\cref{sec:vortexextraction}) and the interactive visualization system (\cref{sec:visualizationsystem}).

\acknowledgments{
We thank the anonymous reviewers for their valuable feedback. This research was supported by NSF OAC 2102761.}

\bibliographystyle{abbrv-doi-hyperref}

\bibliography{main}



\end{document}